\pdfoutput=1

\documentclass[a4paper,cits]{JINST}
\usepackage{microtype}
\usepackage{booktabs}
\usepackage{amsmath}
\usepackage{lineno}

\newcommand{\bbonu}{\ensuremath{\beta\beta0\nu}}
\newcommand{\Qbb}{\ensuremath{Q_{\beta\beta}}}
\newcommand{\NA}{\ensuremath{^{22}}Na}

\newcommand{\XE}{\ensuremath{{}^{136}\rm Xe}}
\author{
\mbox{The NEXT Collaboration}

V.~\'Alvarez,$^{a}$
F.I.G.~Borges,$^{b}$
S.~C\'arcel,$^{a}$
J.~Castel,$^{c}$
S.~Cebri\'an,$^{c}$
A.~Cervera,$^{a}$
C.A.N.~Conde,$^{b}$
T.~Dafni,$^{c}$
T.H.V.T.~Dias,$^{b}$
J.~D\'iaz,$^{a}$
M.~Egorov,$^{d}$
R.~Esteve,$^{e}$
P.~Evtoukhovitch,$^{f}$
L.M.P.~Fernandes,$^{b}$
P.~Ferrario,$^{a}$
A.L.~Ferreira,$^{g}$
E.D.C.~Freitas,$^{b}$
V.M.~Gehman,$^{d}$
A.~Gil,$^{a}$
A.~Goldschmidt,$^{d}$
H.~G\'omez,$^{c}$
J.J.~G\'omez-Cadenas,$^{a}$\thanks{Spokesperson (gomez@mail.cern.ch)} ~
D.~Gonz\'alez-D\'iaz,$^{c}$
R.M.~Guti\'errez,$^{h}$
J.~Hauptman,$^{i}$
J.A.~Hernando Morata,$^{j}$
D.C.~Herrera,$^{c}$
F.J.~Iguaz,$^{c}$
I.G.~Irastorza,$^{c}$
M.A.~Jinete,$^{h}$
L.~Labarga,$^{k}$
A.~Laing,$^{a}$
I.~Liubarsky,$^{a}$
J.A.M.~Lopes,$^{b}$
D.~Lorca,$^{a}$
M.~Losada,$^{h}$
G.~Luz\'on,$^{c}$
A.~Mar\'i,$^{e}$
J.~Mart\'in-Albo,$^{a}$\thanks{Co-corresponding author} ~
A.~Mart\'inez,$^{a}$
T.~Miller,$^{d}$
A.~Moiseenko,$^{f}$
F.~Monrabal,$^{a}$\thanks{Co-corresponding author} ~
C.M.B.~Monteiro,$^{b}$
F.J.~Mora,$^{e}$
L.M. Moutinho,$^{g}$
J.~Mu\~noz~Vidal,$^{a}$
H.~Natal da Luz,$^{b}$
G.~Navarro,$^{h}$
M.~Nebot-Guinot,$^{a}$
D.~Nygren,$^{d}$
C.A.B.~Oliveira,$^{d}$
R.~Palma,$^{l}$
J.~P\'erez,$^{k}$
J.L.~P\'erez~Aparicio,$^{l}$
J.~Renner,$^{d}$
L.~Ripoll,$^{m}$
A.~Rodr\'iguez,$^{c}$
J.~Rodr\'iguez,$^{a}$
F.P.~Santos,$^{b}$
J.M.F.~dos~Santos,$^{b}$
L.~Segu\'i,$^{c}$
L.~Serra,$^{a}$
D.~Shuman,$^{d}$
A.~Sim\'on,$^{a}$
C.~Sofka,$^{n}$
M.~Sorel,$^{a}$
J.F.~Toledo,$^{d}$
A.~Tom\'as,$^{c}$
J.~Torrent,$^{m}$
Z.~Tsamalaidze,$^{f}$
D.~V\'azquez,$^{j}$
J.F.C.A.~Veloso,$^{g}$
J.A.~Villar,$^{c}$
R.~Webb,$^{n}$
J.T.~White,$^{n}$
N.~Yahlali$^{a}$\\
\llap{$^{a}$}
Instituto de F\'isica Corpuscular (IFIC), CSIC \& Universitat de Val\`encia\\
Calle Catedr\'atico Jos\'e Beltr\'an, 2, 46980 Paterna, Valencia, Spain\\
\llap{$^{b}$}
Departamento de Fisica, Universidade de Coimbra\\
Rua Larga, 3004-516 Coimbra, Portugal\\
\llap{$^c$}
Lab.\ de F\'isica Nuclear y Astropart\'iculas, Universidad de Zaragoza\\ 
Calle Pedro Cerbuna, 12, 50009 Zaragoza, Spain\\
\llap{$^d$}
Lawrence Berkeley National Laboratory (LBNL)\\
1 Cyclotron Road, Berkeley, California 94720, USA\\
\llap{$^{e}$}
Instituto de Instrumentaci\'on para Imagen Molecular (I3M), Universitat Polit\`ecnica de Val\`encia\\ 
Camino de Vera, s/n, Edificio 8B, 46022 Valencia, Spain\\
\llap{$^{f}$}
Joint Institute for Nuclear Research (JINR)\\
Joliot-Curie 6, 141980 Dubna, Russia\\
\llap{$^{g}$}Institute of Nanostructures, Nanomodelling and Nanofabrication (i3N), Universidade de Aveiro\\
Campus de Santiago, 3810-193 Aveiro, Portugal\\
\llap{$^{h}$}
Centro de Investigaciones, Universidad Antonio Nari\~no\\ 
Carretera 3 este No.\ 47A-15, Bogot\'a, Colombia\\
\llap{$^{i}$}
Department of Physics and Astronomy, Iowa State University\\
12 Physics Hall, Ames, Iowa 50011-3160, USA\\
\llap{$^{j}$}
Instituto Gallego de F\'isica de Altas Energ\'ias (IGFAE), Univ.\ de Santiago de Compostela\\
Campus sur, R\'ua Xos\'e Mar\'ia Su\'arez N\'u\~nez, s/n, 15782 Santiago de Compostela, Spain\\
\llap{$^{k}$}
Departamento de F\'isica Te\'orica, Universidad Aut\'onoma de Madrid\\
Ciudad Universitaria de Cantoblanco, 28049 Madrid, Spain\\
\llap{$^{l}$}
Dpto.\ de Mec\'anica de Medios Continuos y Teor\'ia de Estructuras, Univ.\ Polit\`ecnica de Val\`encia\\
Camino de Vera, s/n, 46071 Valencia, Spain\\
\llap{$^{m}$}
Escola Polit\`ecnica Superior, Universitat de Girona\\
Av.~Montilivi, s/n, 17071 Girona, Spain\\
\llap{$^{n}$}
Department of Physics and Astronomy, Texas A\&M University\\
College Station, Texas 77843-4242, USA\\

E-mail: \email{justo.martin-albo@ific.uv.es, francesc.monrabal@ific.uv.es}
}
\title{Initial results of NEXT-DEMO, a large-scale prototype of the NEXT-100 experiment}

\abstract{
NEXT-DEMO is a large-scale prototype of the NEXT-100 detector, an electroluminescent time projection chamber that will search for the neutrinoless double beta decay of \XE\ using 100--150 kg of enriched xenon gas. NEXT-DEMO was built to prove the expected performance of NEXT-100, namely, energy resolution better than 1\% FWHM at 2.5 MeV and event topological reconstruction. In this paper we describe the prototype and its initial results. A resolution of 1.75\% FWHM at 511 keV (which extrapolates to 0.8\% FWHM at 2.5 MeV) was obtained at 10 bar pressure using a gamma-ray calibration source. Also, a basic study of the event topology along the longitudinal coordinate is presented, proving that it is possible to identify the distinct d$E/$d$x$ of electron tracks in high-pressure xenon using an electroluminescence TPC.}

\preprint{arXiv:1211.4838}

\begin{document}

\tableofcontents

%%% SECTION 1. INTRODUCTION
%%%%%%%%%%%%%%%%%%%%%%%%%%%%%%%%%%%%%%%%%%%%%%%%%%%%%%%%%%%%
\section{Introduction} \label{sec:Introduction}
%%%
The NEXT-100 \cite{Alvarez:2012haa} time projection chamber (TPC), currently under construction, will search for neutrinoless double beta decay (\bbonu) using 100--150 kg of xenon gas enriched in the \XE\ isotope to $91\%$. The detector boasts two important features for \bbonu\ searches: \emph{excellent energy resolution} (better than 1\% FWHM at the $Q$ value of \XE) and event \emph{topological information} for the identification of signal and background. This combination gives NEXT an excellent experimental sensitivity to \bbonu\ \cite{GomezCadenas:2011it}. In addition, the technique can be extrapolated to the ton-scale, thus allowing the full exploration of the inverted hierarchy of neutrino masses \cite{GomezCadenas:2012jv}.

Xenon, as a detection medium, provides both \emph{scintillation} and \emph{ionization} as primary signals. The former is used in NEXT to establish the start-of-event time ($t_{0}$), while the latter is used for calorimetry and tracking. In its gaseous phase, xenon can provide high energy resolution, in principle as good as 0.3\% FHWM at 2.5~MeV \cite{Nygren:2009zz}. In order to achieve optimal energy resolution, the ionization signal is amplified in NEXT using the \emph{electroluminescence} (EL) of xenon: the electrons liberated by ionizing particles passing through the gas are first drifted towards the TPC anode by a weak electric field ($\sim300~\mathrm{V/cm}$), entering then into another region where they are accelerated by a high electric field ($\sim25~\mathrm{kV/cm}$ at 10 bar), intense enough so that the electrons can excite the xenon atoms but not enough to ionize them. This excitation energy is ultimately released in the form of proportional (with sub-poissonian fluctuations) \emph{secondary scintillation} light.

%%%%%%%%%%
\begin{figure}
\centering
\includegraphics[width=0.85\textwidth]{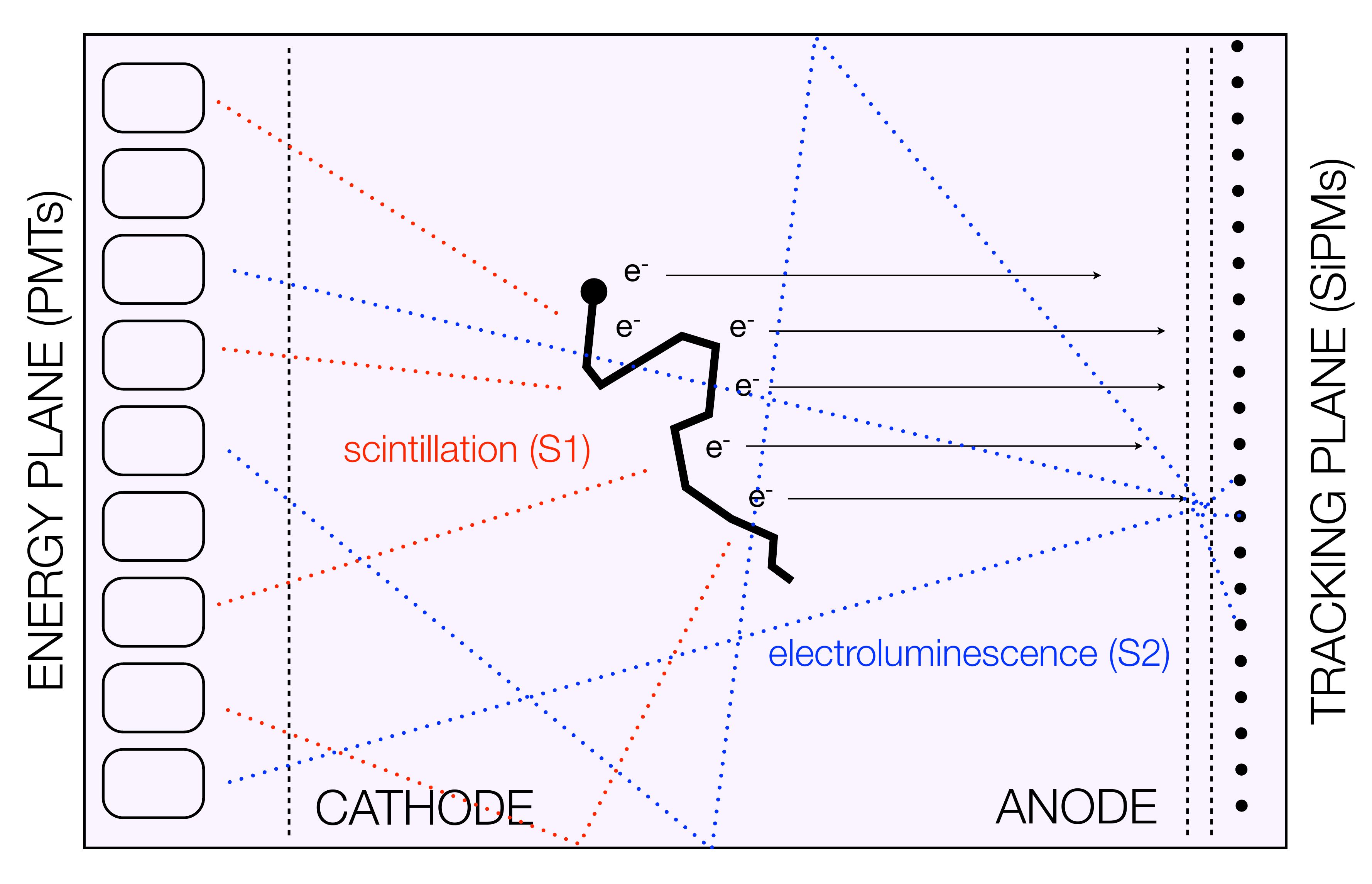}
\caption{The \emph{Separated, Optimized Functions} (SOFT) concept in the NEXT experiment: EL light generated at the anode is recorded in the photosensor plane right behind it and used for tracking; it is also recorded in the photosensor plane behind the transparent cathode and used for a precise energy measurement.} \label{fig:SOFT}
\end{figure}
%%%%%%%%%%

NEXT-100 will have different readout systems for calorimetry and tracking (see figure \ref{fig:SOFT}). An array of photomultiplier tubes (PMTs), the so-called \emph{energy plane}, located behind the TPC cathode detects a fraction of the secondary scintillation (S2) light to provide a precise measurement of the total energy deposited in the gas. These PMTs detect as well the primary scintillation (S1), used to signal the start of the event. The forward-going S2 light is detected by a dense array of silicon photomultipliers (SiPMs), known as the \emph{tracking plane}, located behind the anode, very close to the EL region, and is used for event topological reconstruction.

During the last three years, as part of the NEXT R\&D program, we have built the NEXT-DEMO prototype with the double aim of demonstrating the NEXT detector concept and gaining experience that would facilitate the design, construction and operation of the NEXT-100 detector. This paper describes the prototype and presents the results of the initial operation. The organization is as follows: section~\ref{sec:Detector} describes in detail the apparatus; section~\ref{sec:Na22} explains its characterization with a \NA\ source; section~\ref{sec:EnergyResolution} presents our initial energy resolution analysis, and section~\ref{sec:Topology} first results on topology reconstruction; section~\ref{sec:Summary} concludes.

%%% SECTION 2. THE NEXT-DEMO PROTOTYPE
%%%%%%%%%%%%%%%%%%%%%%%%%%%%%%%%%%%%%%%%%%%%%%%%%%%%%%%%%%%%
\section{The NEXT-DEMO prototype} \label{sec:Detector}
\begin{figure}
\centering
\includegraphics[width=0.75\textwidth]{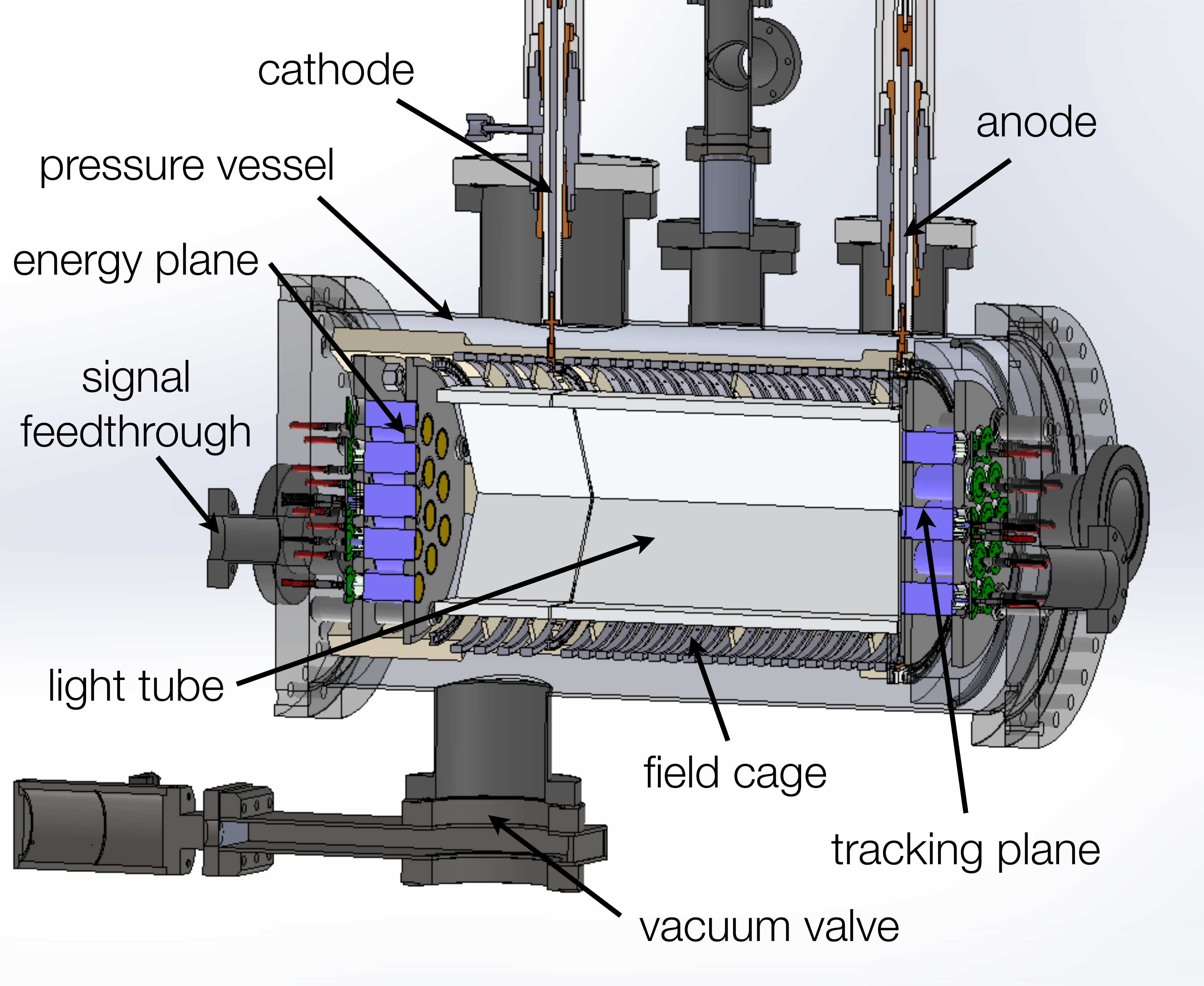}
\caption{Cross-section drawing of the NEXT-DEMO detector with all major parts labelled.} \label{fig:NextDemoXSec}
\end{figure}

\begin{figure}
\centering
\includegraphics[width=\textwidth]{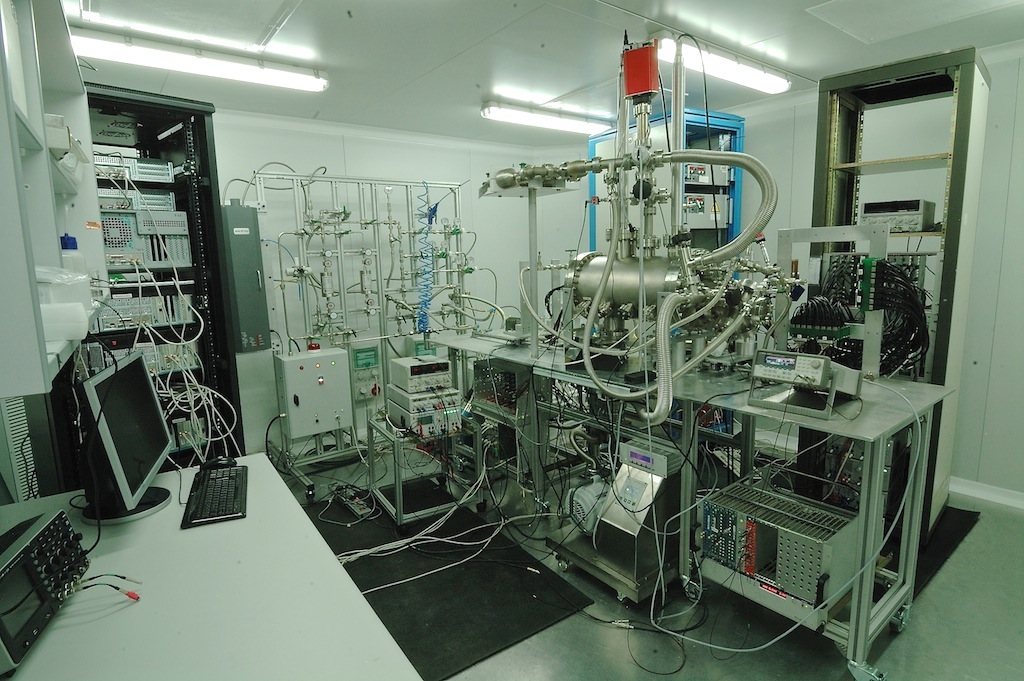}
\caption{The NEXT-DEMO detector and ancillary systems (gas system, front-end electronics and DAQ) in their location at a semi-clean room at IFIC.} 
\label{fig:NextDemoCleanRoom}
\end{figure}

%%%
NEXT-DEMO, shown in figure~\ref{fig:NextDemoXSec}, is a high-pressure xenon electroluminescent TPC implementing the NEXT detector concept described above. Its active volume is 30 cm long and 30 cm diameter. A tube of hexagonal cross section made of PTFE is inserted into the active volume to improve the light collection. The TPC is housed in a stainless-steel pressure vessel that can withstand up to 15~bar. Natural xenon circulates in a closed loop through the vessel and a system of purifying filters. The detector is not radiopure and is not shielded against natural radioactivity. It is installed in a semi-clean room (see figure~\ref{fig:NextDemoCleanRoom}) at the \emph{Instituto de F\'isica Corpuscular} (IFIC), in Valencia, Spain.

%%%%%%%%%%
%%%%%%%%%%

%%%%%%%%%%
%%%%%%%%%%

The main objective of NEXT-DEMO was the validation of the NEXT-100 design. More specifically, the goals of the prototype were the following: (a) to demonstrate good energy resolution in a large active volume; (b) to reconstruct the topological signature of electrons in high-pressure xenon gas (HPXe); (c) to test long drift lengths and high voltages; (d) to understand gas recirculation and purification in a large volume, including operation stability and robustness against leaks; and (e) to understand the collection of light and the use of wavelength shifters (WLS).

This paper describes the initial operation of NEXT-DEMO with a tracking plane implemented using 19 pressure-resistant photomultipliers, identical to those used in the energy plane but operated at a lower gain. Instrumenting the tracking plane with PMTs --- unlike NEXT-100, which will use SiPMs --- during this period simplified the initial commissioning, debugging and operation of the detector due to the smaller number of readout channels (19 PMTs in contrast to the 256 SiPMs projected for the second phase of NEXT-DEMO) and their intrinsic sensitivity to the UV light emitted by xenon.

The detector response was studied under two different conditions: an \emph{ultraviolet configuration} (UVC) in which the PTFE \emph{light tube} had no coating, and a \emph{blue configuration} (BC) in which the panels were coated with \emph{tetraphenyl butadiene} (TPB), a wavelength shifter, in order to study the possible improvement in light collection.

%%%%%%%%%%%%%%%%%%%%%%%%%%%%%%%%%%%%%%%%%%%%%%%%%%%%%%%%%%%%
\subsection{Gas system} \label{subsec:GasSystem}

\begin{figure}
\centering
\includegraphics[width=\textwidth]{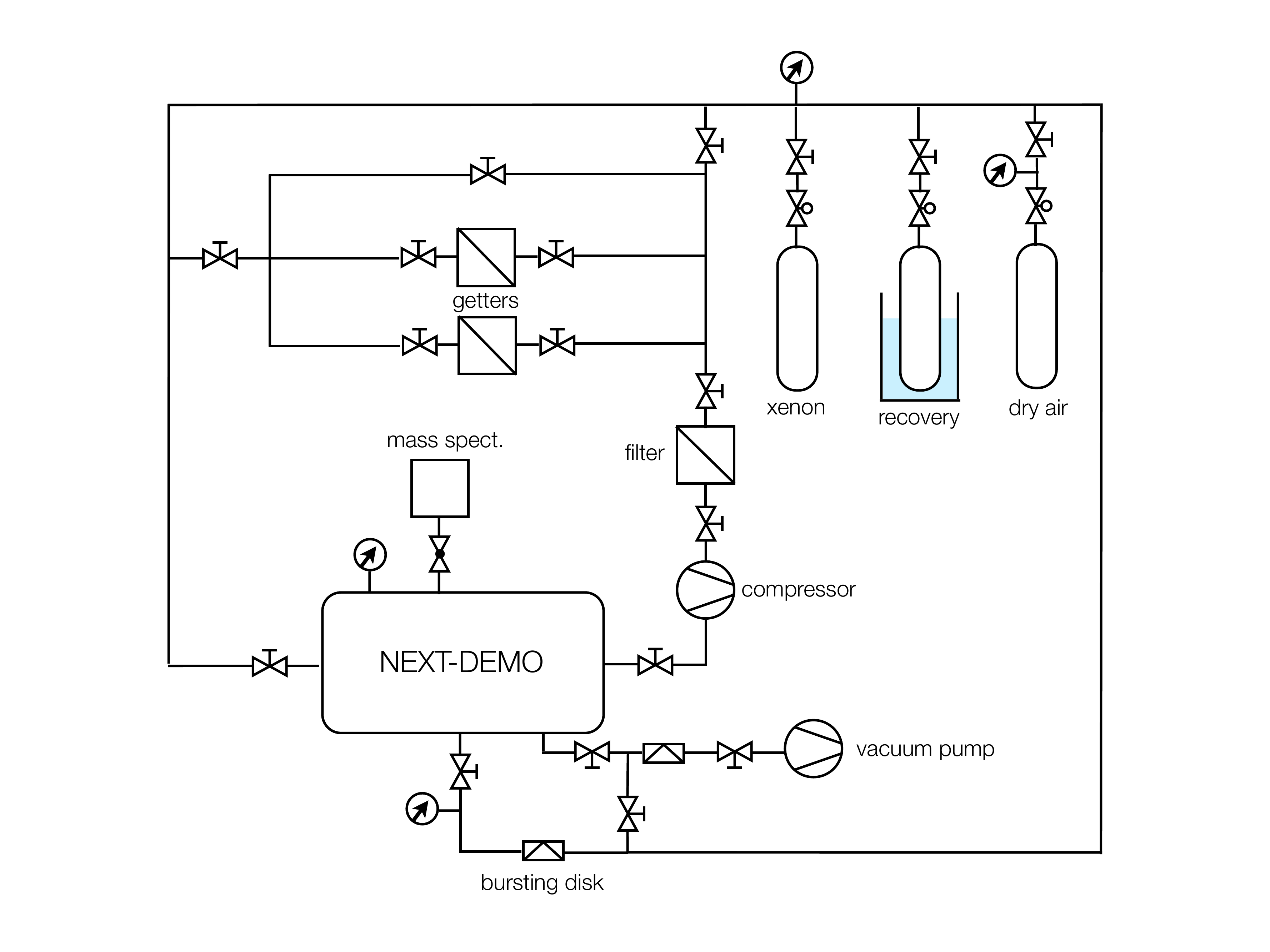}
\caption{Simplified schematic of the gas system of NEXT-DEMO.} \label{fig:GasSystem}
\end{figure}

The functions of the gas system of NEXT-DEMO are the evacuation of the detector, its pressurization and depressurization with xenon (and argon), and the recirculation of the gas through purification filters. A schematic of the system is shown in figure~\ref{fig:GasSystem}.

%%%%%%%%%%
%%%%%%%%%%

The standard procedure during normal operation of the detector starts with the evacuation of the vessel to vacuum levels around $10^{-5}$~mbar, followed by an argon purge. A second vacuum step exhausts the argon from the system. The detector is then filled with xenon gas to pressures up to 15 bar. The xenon can be cryogenically reclaimed to a stainless-steel bottle connected to the gas system by simply immersing this in a dewar filled with liquid nitrogen. The pressure regulator of the bottle is fully opened to allow the xenon gas to flow inside it (due to the temperature difference) and freeze.

The vacuum pumping system consists of a roughing pump (Edwards XDS5 scroll vacuum pump) and a turbo molecular pump (Pfeiffer HiPace 300). Vacuum pressures better than $10^{-7}$~mbar have been obtained after pumping out the detector for several days. The recirculation loop is powered by an oil-less, single-diaphragm compressor (KNF PJ24999-2400) with a nominal flow of 100 standard liters per minute. This translates to 
an approximate flow of 10 liters per minute at 10 bar, thus recirculating the full volume of NEXT-DEMO ($\sim45$~L) in about 5 minutes. The gas system is equipped with both room-temperature (SAES MC50) and heated \emph{getters} (SAES PS4-MT15) that remove electronegative impurities (O$_{2}$, H$_{2}$O, etc.) from the xenon. All the gas piping, save for the inlet gas hoses and getter fittings, are $1/2$ inch diameter with VCR fittings. A set of pressure relief valves (with different settings for the various parts of the system) and a burst disk in the vacuum system protect the equipment and personnel from overpressure hazards.

The operation of the gas system has been, in general, very stable. The detector has run without interruption for long periods of up to 6 weeks with no leaks and continuous purification of the gas. However, one major leak occurred when the diaphragm of the recirculation pump broke, causing the loss of the xenon volume contained in the chamber. This led to the installation of an emergency mechanism that, in the event of pressure drop, automatically closes those valves connecting the pump to the rest of the gas system. Since installation, only one major failure has taken place which the emergency system isolated without loss of gas.  Micro-leaks, on the level of 0.005 bar per day, due to bad connections in the gas system have also been detected making it necessary to introduce additional xenon to maintain pressure.

Several improvements to the initial design of the NEXT-100 gas system have been made thanks to the initial data runs described in this paper. These include the recognition of the importance of reliability of the main pump as well as the decision to use hot getters as the main gas purification stage due to their negligible emission of radon compared to that of room-temperature getters.

%%%%%%%%%%%%%%%%%%%%%%%%%%%%%%%%%%%%%%%%%%%%%%%%%%%%%%%%%%%%
\subsection{Pressure vessel} \label{subsec:PressureVessel}
\begin{figure}
\centering
\includegraphics[scale=0.9]{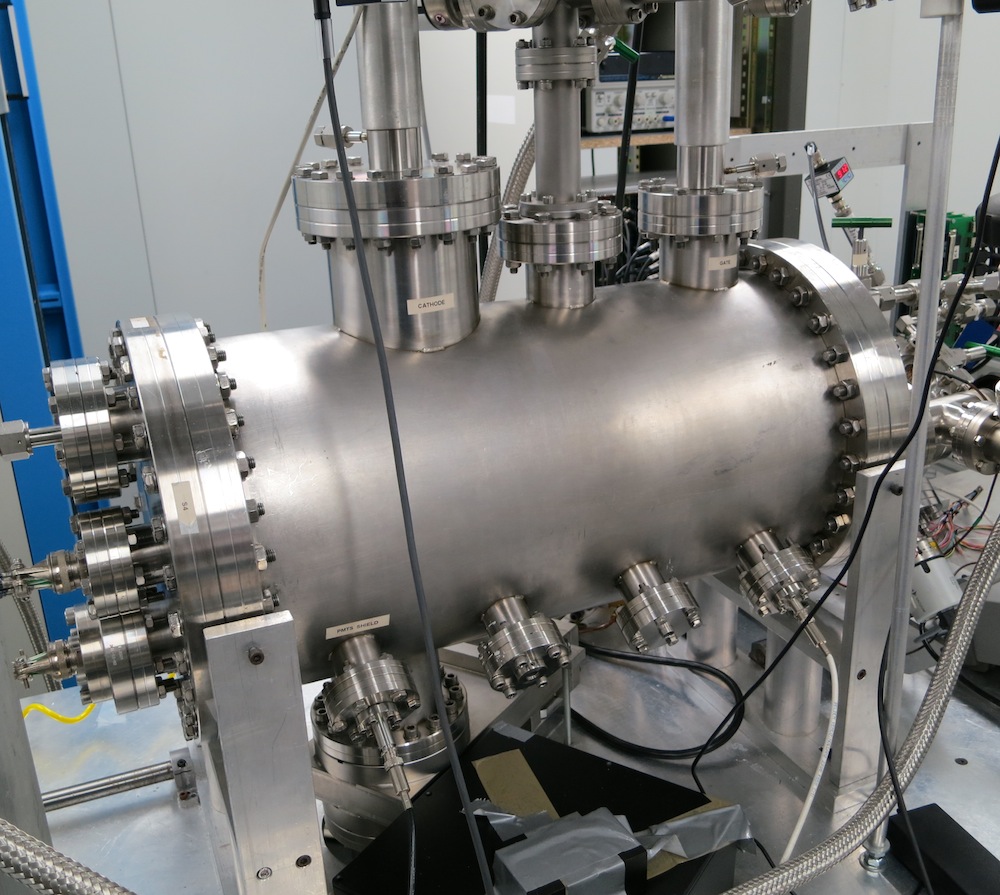}
\caption{The pressure vessel of NEXT-DEMO.} \label{fig:Vessel}
\end{figure}

%%%
The pressure vessel of NEXT-DEMO, shown in figure~\ref{fig:Vessel}, is a stainless-steel (grade 304L) cylindrical shell, 3 mm thick, 30 cm diameter and 60 cm length, welded to CF flanges on both ends. The two end-caps are 3-cm thick plates with standard CF knife-edge flanges. Flat copper gaskets are used as sealing. The vessel was certified to 15 bar operational pressure. It was designed at IFIC and built by Trinos Vacuum Systems, a local manufacturer. Additional improvements --- including the support structure and a rail system to open and move the end-caps --- have been made using the mechanical workshop at IFIC.

%%%%%%%%%%
%%%%%%%%%%

The side of the chamber includes 8 CF40 half-nipples. One set of 4 is located in the horizontal plane while the other is displaced towards the underside with respect to the first set by $60^\circ$. These contain radioactive source ports used for calibration of the TPC. The ports are made by welding a 0.5 mm blank at the end of a 12 mm liquid feedthrough. On top of the vessel and along the vertical plane there are three additional half-nipples (CF130, CF67 and CF80) used for high-voltage feeding and connection to a mass spectrometer (through a leak valve). On the opposite side, at the bottom, a CF100 port connects the pressure vessel to the vacuum pumping system. A guillotine valve closes this connection when the vessel is under pressure. The end-caps include several CF ports for the connections to the gas recirculation loop and for the feedthroughs (power and signal) of the PMT planes.

%%%%%%%%%%%%%%%%%%%%%%%%%%%%%%%%%%%%%%%%%%%%%%%%%%%%%%%%%%%%
\subsection{Time projection chamber} \label{subsec:TPC}
\begin{figure}
\centering
\includegraphics[width=0.75\textwidth]{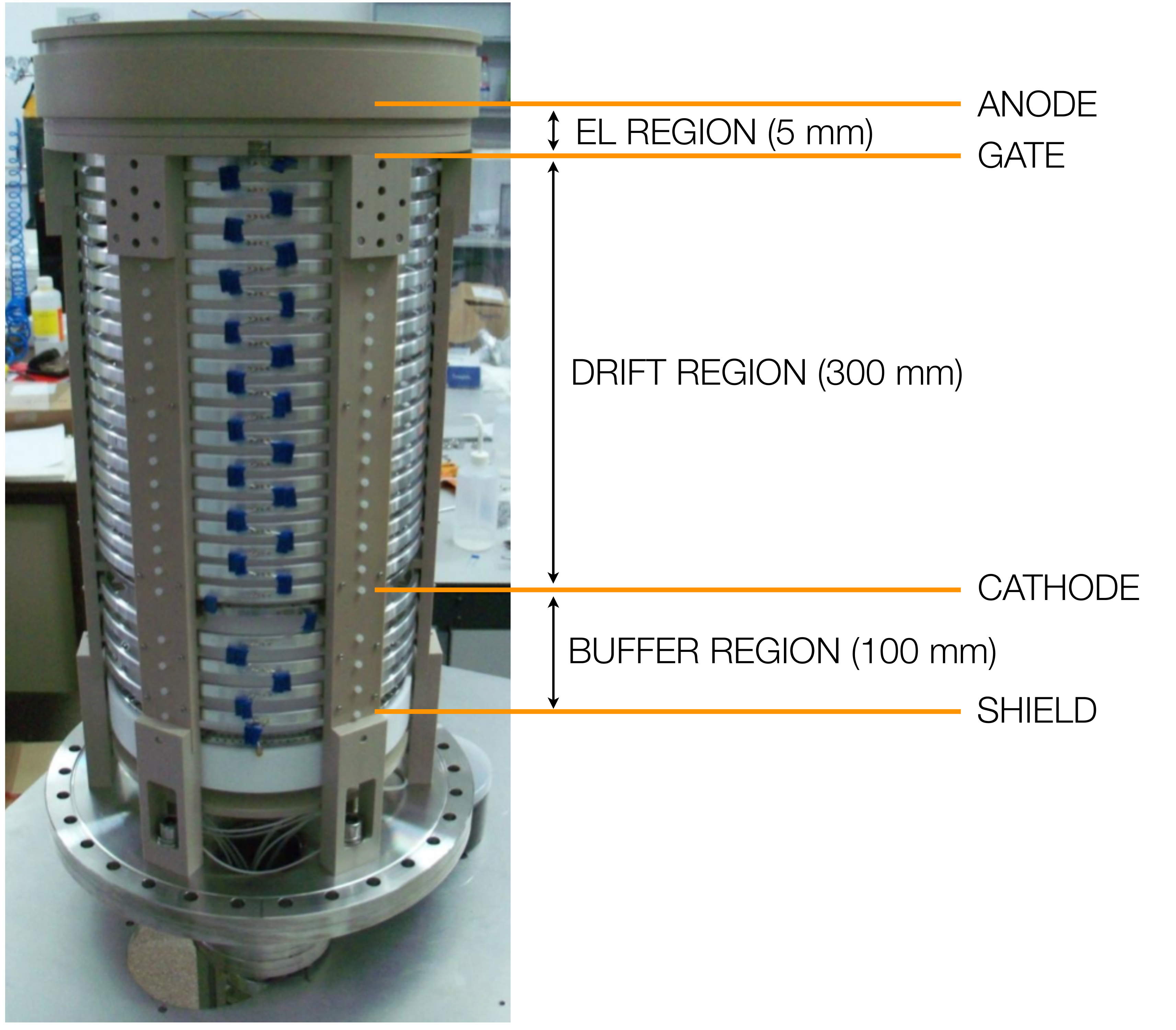}
\caption{External view of the time projection chamber mounted on one end-cap. The approximate positions of the different regions of the TPC are indicated.} \label{fig:TPC}
\end{figure}

%%%
Three metallic wire grids --- referred to as \emph{cathode}, \emph{gate} and \emph{anode} --- define the two active regions of the chamber (see figure~\ref{fig:TPC}): the 30 cm long \emph{drift region}, between cathode and gate; and the 0.5 cm long \emph{EL region}, between gate and anode. Gate and anode were built using stainless-steel meshes with 88\% open area (30-$\mu$m diameter wires, 50 wires$/$inch) clamped in a tongue-and-groove circular frame with a tensioning ring that is torqued with set screws to achieve the optimum tension. The cathode was built in a similar fashion by clamping parallel wires 1~cm apart into another circular frame.  

%%%%%%%%%%
%%%%%%%%%%

The electric field in the TPC is created by supplying a large negative voltage to the cathode then degrading it across the drift region using a series of metallic rings of 30 cm diameter spaced 5 mm apart and connected via 0.5~G$\Omega$ resistors. The rings were manufactured by cutting and machining aluminum pipe. The gate is at negative voltage so that a moderate electric field --- typically of 2.5 to 3 $\mathrm{kV~cm^{-1}~bar^{-1}}$ --- is created between the gate and the anode, which is at ground.
A \emph{buffer region} of 10 cm between the cathode and the energy plane protects this from the high-voltage by degrading it safely to ground potential.

\begin{figure}
\centering
\includegraphics[scale=.7]{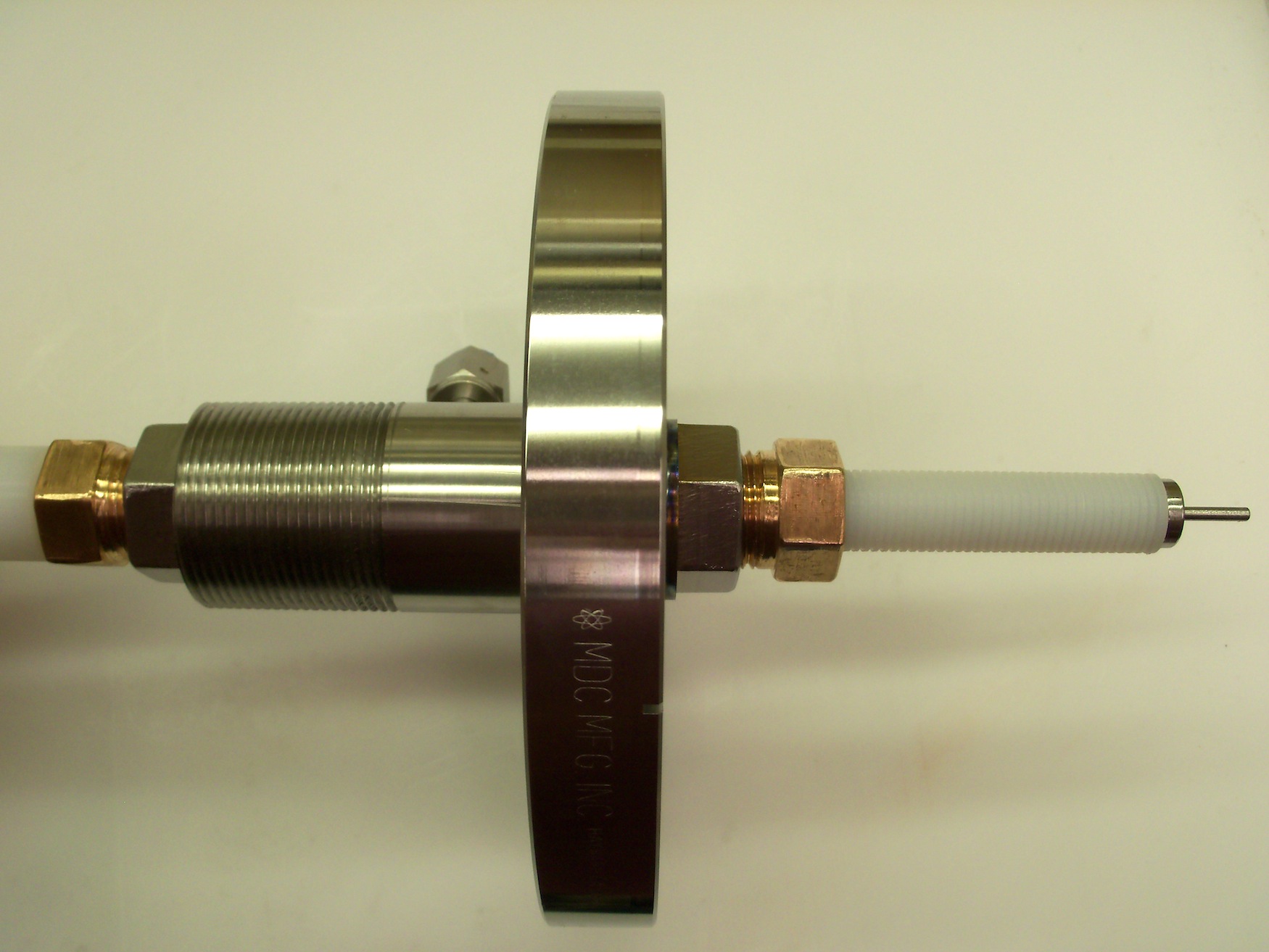} 
\caption{The NEXT-DEMO high-voltage feed-through, designed and built by Texas A\&M.}
\label{fig:HVFT}
\end{figure}

The high voltage is supplied to the cathode and the gate through custom-made high-voltage feed-throughs (HVFT), shown in figure~\ref{fig:HVFT}, built pressing a stainless-steel rod into a Tefzel (a plastic with high dielectric strength) tube, which is then clamped using plastic ferrules to a CF flange. They have been tested to high vacuum and 100 kV without leaking or sparking. 

%%%%%%%%%%
%%%%%%%%%%

\begin{figure}
\centering
\includegraphics[height=6.75cm]{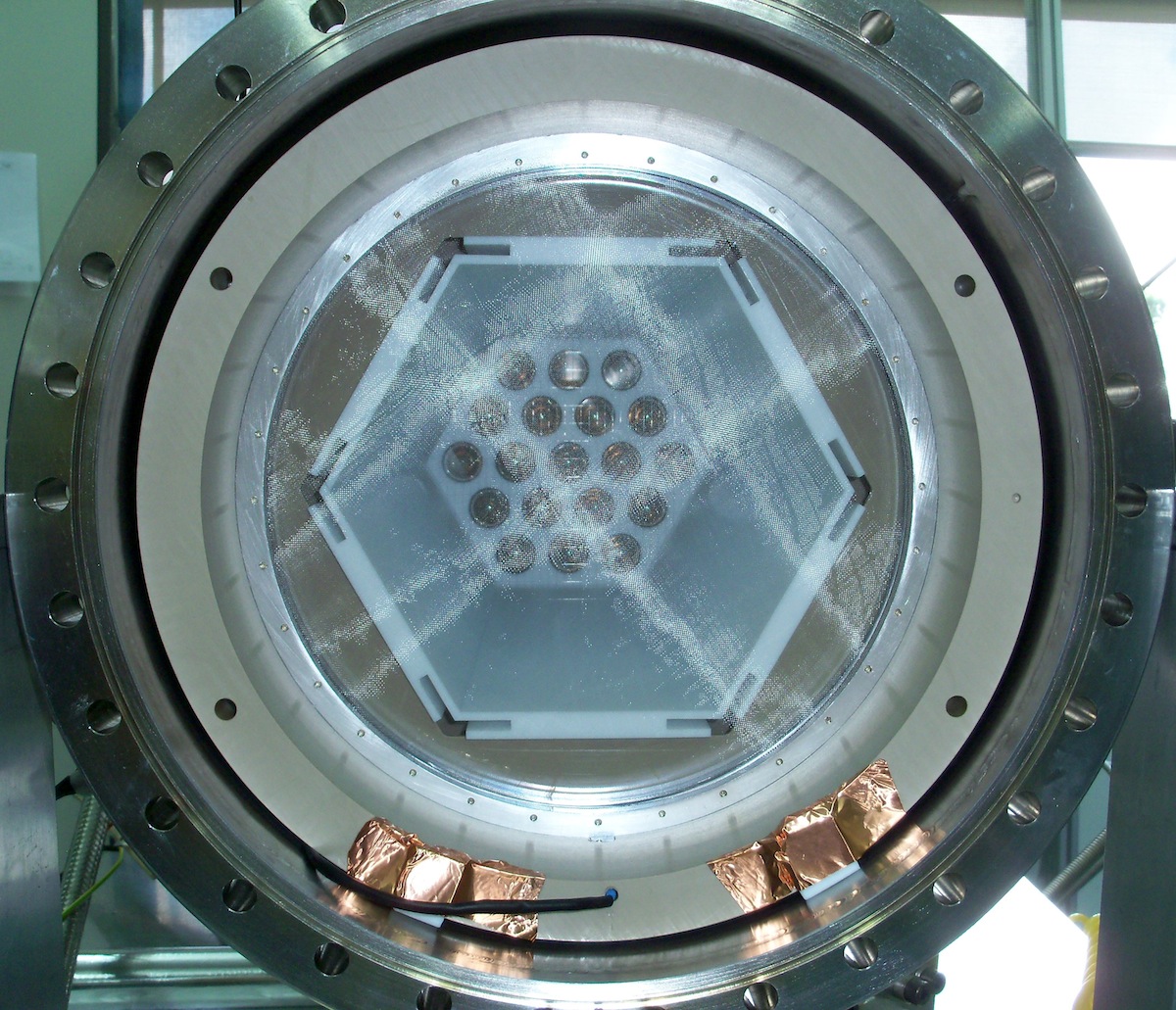}
\includegraphics[height=6.75cm]{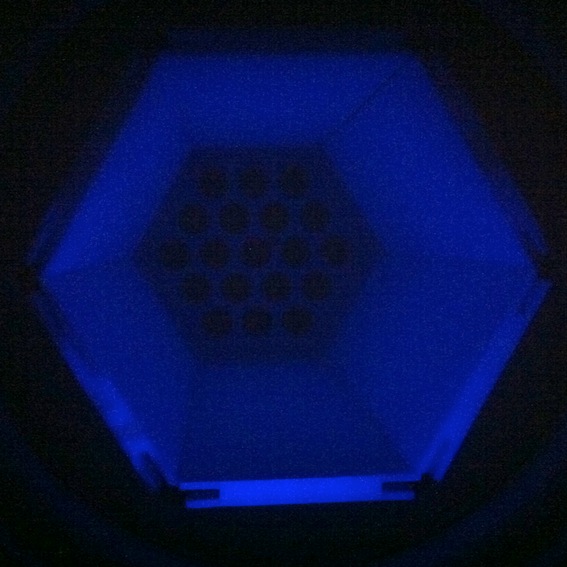}  
\caption{View of the light tube from the position of the tracking plane. Left: The meshes of the EL region can be seen in the foreground, and in the background, at the end of the light tube, the PMTs of the energy plane are visible. Right: The light tube of NEXT-DEMO illuminated with a UV lamp after being coated with TPB.} \label{fig:LightTube}
\end{figure}

A set of six panels made of PTFE (Teflon) are mounted inside the electric-field cage forming a \emph{light tube} of hexagonal cross section (see figure~\ref{fig:LightTube}) with and apothem length of 8 cm. PTFE is known to be an excellent reflector in a wide range of wavelengths \cite{Silva:2009ip}, thus improving the light collection efficiency of the detector. In a second stage, the panels were vacuum-evaporated with TPB --- which shifts the UV light emitted by xenon to blue ($\sim430$~nm) --- in order to study the  improvement in reflectivity and light detection. Figure~\ref{fig:LightTube} (right panel) shows the light tube illuminated with a UV lamp after the coating.

%%%%%%%%%%
%%%%%%%%%%

Six bars manufactured from PEEK, a low outgassing plastic, hold the electric-field cage and the energy plane together. The whole structure is attached to one of the end-caps using screws, and introduced inside the vessel with the help of a rail system. All the TPC structures and the HVFT were designed and built by Texas A\&M.

%%%%%%%%%%%%%%%%%%%%%%%%%%%%%%%%%%%%%%%%%%%%%%%%%%%%%%%%%%%%
\subsection{Detection planes} \label{subsec:DetPlanes}
\begin{figure}
\centering
\includegraphics[height=6.5cm]{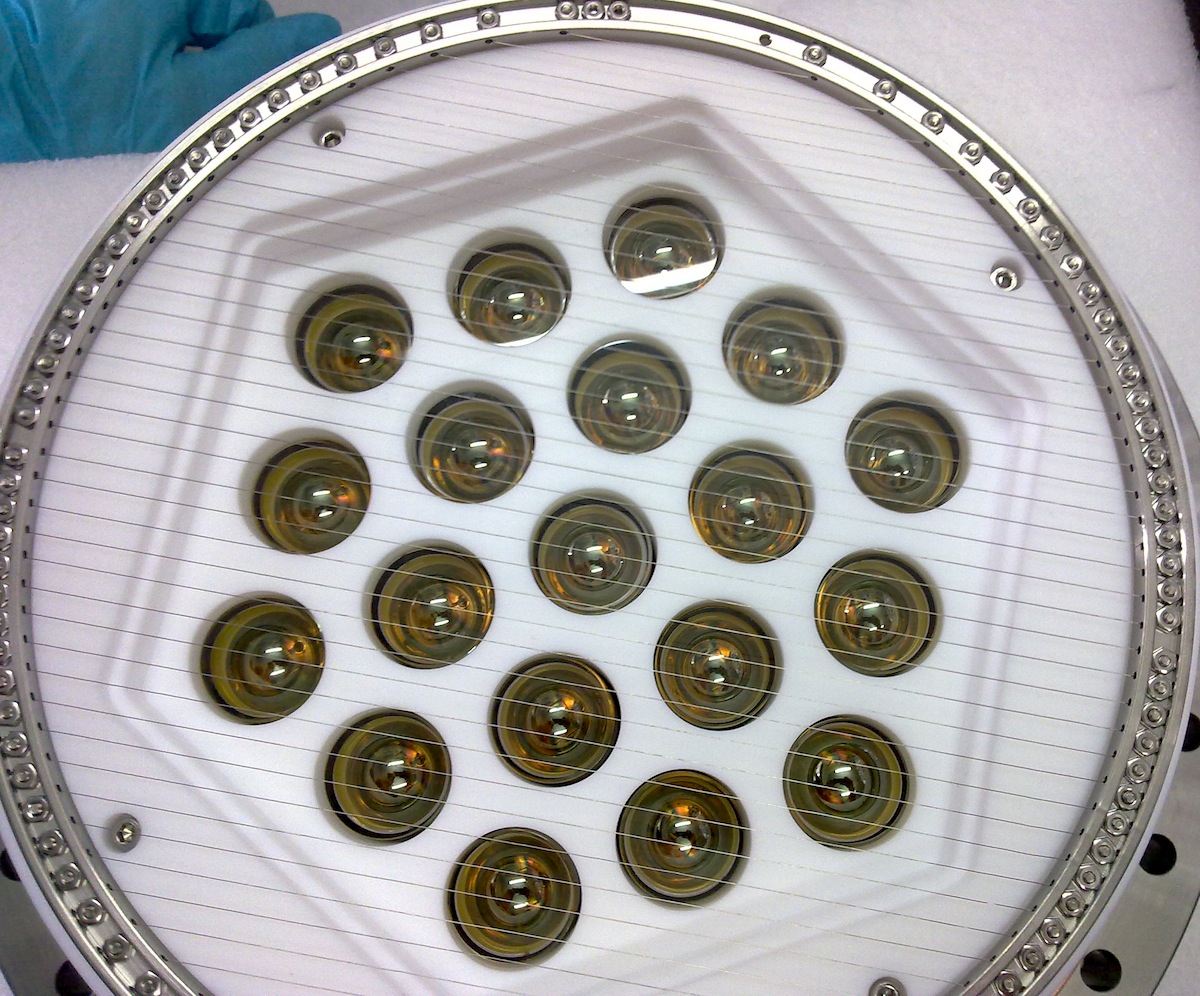}
\includegraphics[height=6.5cm]{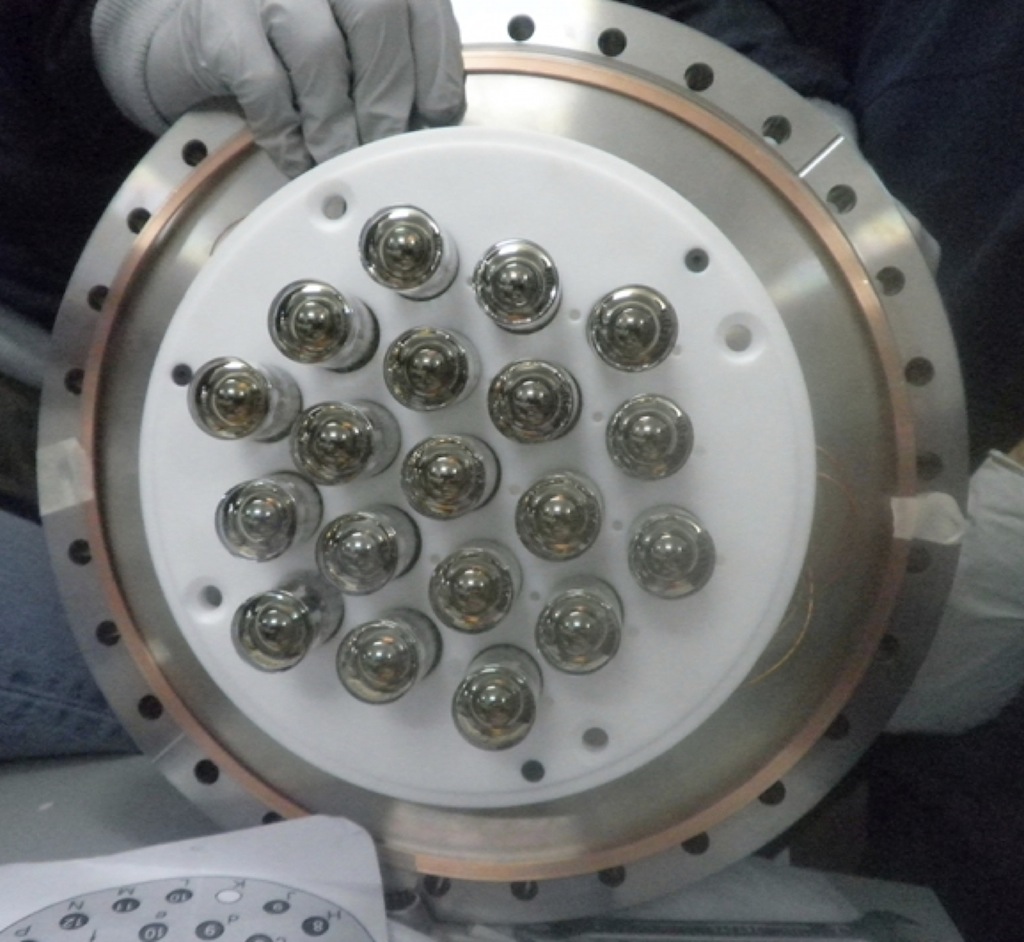}
\caption{The energy (left) and tracking (right) planes of NEXT-DEMO, each one equipped with 19 Hamamatsu R7378A PMTs.} \label{fig:DetPlanes}
\end{figure}

%%%
In NEXT-DEMO, the energy plane (see figure~\ref{fig:DetPlanes}, left panel) is equipped with 19 Hamamatsu R7378A photomultiplier tubes. These are 1-inch, pressure-resistant (up to 20 bar) PMTs with acceptable quantum efficiency ($\sim 15$\%) in the VUV region. The resulting photocathode coverage of the energy plane is about 39\%. The PMTs are inserted into a PTFE holder following a hexagonal pattern. A grid, known as \emph{shield} and similar to the cathode but with the wires spaced 0.5~cm apart, is screwed on top of the holder and set to electrical ground. As explained above, this protects the PMTs from the high-voltage set in the cathode, and ensures that the electric field in the 10-cm buffer region is below the EL threshold.

The PMTs are connected to custom-made electrical bases that are used as voltage dividers, and also allow the extraction of the signal induced in the PMTs. This requires a total of 38 cables inside the pressure vessel connected via feed-throughs.  

As mentioned already, the first tracking plane of the NEXT-DEMO detector also uses  19 Hamamatsu R7378A PMTs, as shown in figure~\ref{fig:DetPlanes} (right), but operated at lower gain. They are also held by a PTFE honeycomb, mirroring the energy plane. The PMT windows are located 2 mm away from the anode mesh. Position reconstruction is based on energy sharing between the PMTs, being therefore much better than the distance between PMTs (35 mm from center to center).

%%%%%%%%%%
%%%%%%%%%%

%%%%%%%%%%%%%%%%%%%%%%%%%%%%%%%%%%%%%%%%%%%%%%%%%%%%%%%%%%%%
\subsection{Electronics and DAQ} \label{subsec:Electronics}
%%%
The two optical, primary signals in the NEXT detector concept are in very different scales, and the photomultipliers and their front-end electronics must be ready to handle both. Primary scintillation results in weak (a few photoelectrons per photomultiplier) and fast (the bulk of the signal comes in about 20 ns) signals, whereas the secondary scintillation --- that is, the EL-amplified ionization --- is intense (hundreds to thousands of photoelectrons per PMT) and slow (several microseconds long). 

The gain of the PMTs in NEXT-DEMO was adjusted to around $5\times10^6$ for the energy plane to place the mean amplitude of a single photoelectron pulse well above electronic system noise, and approximately half that for the tracking plane since they record the direct secondary-scintillation light produced in the EL region and, as such, would have a higher probability of saturation at the same gain as those of the cathode.

The PMTs produce fast signals (less than 5~ns wide) making necessary the shaping and filtering of the detector output so that they match the sampling rate of the digitizer. This process also performs the important function of eliminating high frequency noise. An integrator is implemented by simply adding a capacitor and a resistor to the PMT base. The charge integration capacitor shunting the anode stretches the pulse and reduces the primary signal peak voltage accordingly. The design uses a single amplification stage based on a fully differential amplifier THS4511, which features low noise ($2~\mathrm{nV/\sqrt{Hz}}$) and provides enough gain to compensate for the attenuation in the following stage. Amplification is followed by a passive RC filter with a cut frequency of 800 kHz. This filtering produces enough signal stretching to allow the acquisition of many samples per single photo-electron at 40 MHz. The front-end circuit for NEXT-DEMO was implemented in 7 channel boards and connected via HDMI cables to 12-bit 40-MHz digitizer cards. These digitizers are read out by the FPGA-based DAQ modules (FEC cards) that buffer, format and send event fragments to the DAQ PCs. As for the FEC card, the 16-channel digitizer add-in card was designed in a joint effort between CERN and the NEXT Collaboration within the RD-51 program \cite{Martoiu:2011}. These two cards are edge mounted to form a standard 6U220 mm Eurocard. An additional FEC module with a different plug-in card is used as trigger module. Besides forwarding a common clock and commands to all the DAQ modules, it receives trigger candidates from the DAQ modules, runs a trigger algorithm in the FPGA and distributes a trigger signal. The trigger electronics also accepts external triggers for detector calibration purposes.

%%% SECTION 3. ANALYSIS OF NA-22 GAMMA-RAY DATA
%%%%%%%%%%%%%%%%%%%%%%%%%%%%%%%%%%%%%%%%%%%%%%%%%%%%%%%%%%%%
\section{Analysis of \NA\ gamma-ray data} \label{sec:Na22}
\begin{figure}
\centering
\includegraphics[width=0.65\textwidth]{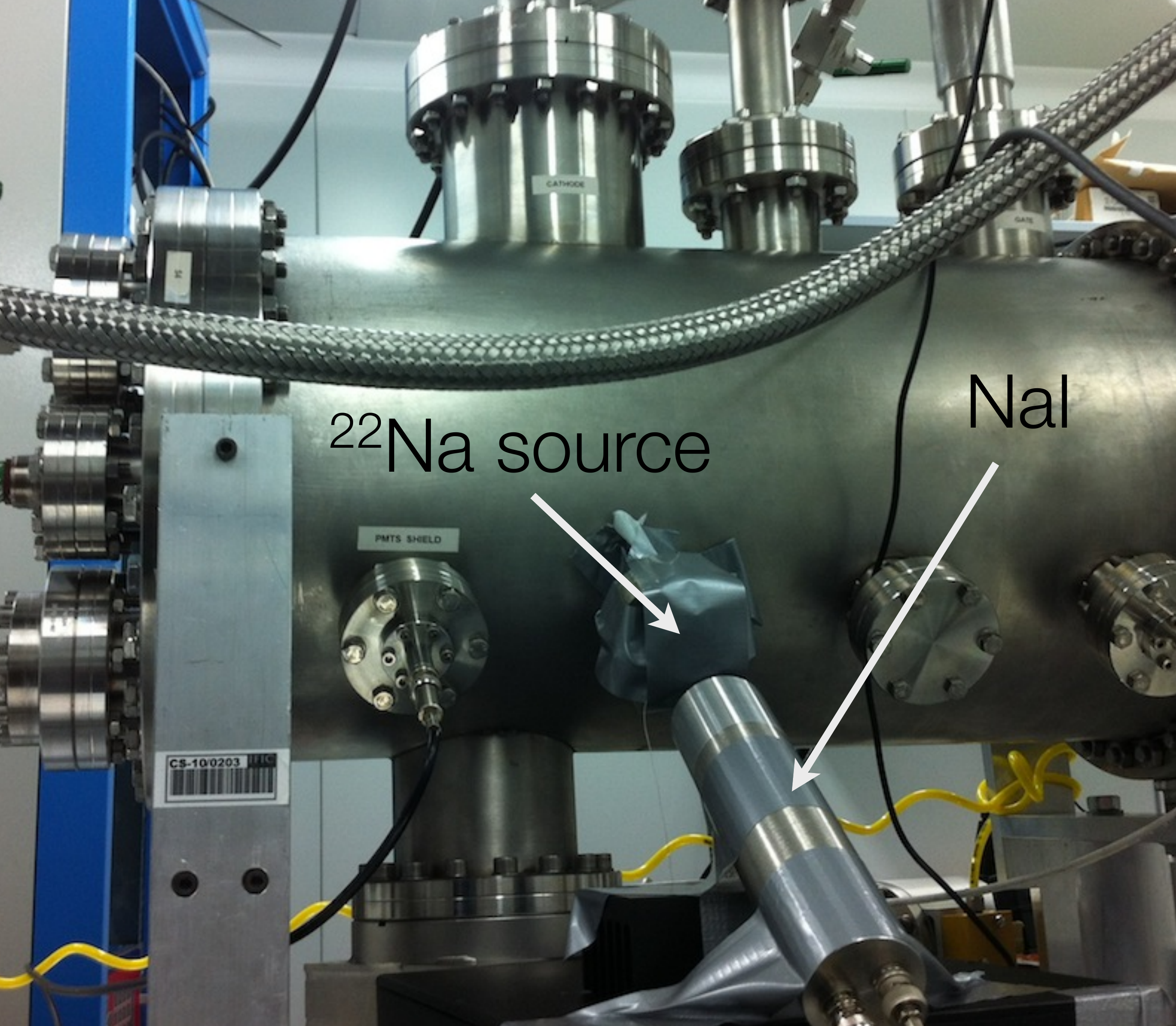}
\caption{A view of NEXT-DEMO with the \NA\ calibration source installed in the port closest to the TPC cathode with the NaI scintillator used for event tagging visible in the foreground.} 
\label{fig:TaggingNaI}
\end{figure}

%%%
The response of the NEXT-DEMO detector was studied using a 1-$\mu$Ci \NA\ calibration source placed at the port near the TPC cathode. Sodium-22 is a long-lived $\beta^{+}$ radioactive isotope, and the annihilation of the positron emitted in its decay, which rarely travels more than 1~mm, results in two back-to-back 511-keV gammas. The coincident detection of the forward gamma in the TPC and of the backward gamma in an external NaI scintillator coupled to a photomultiplier (see figure~\ref{fig:TaggingNaI}) was used to trigger detector read-out. This arrangement optimized the acquisition of useful calibration data.

%%%%%%%%%%
%%%%%%%%%%

Two data sets are discussed in this paper --- one for the \emph{ultraviolet configuration} (UVC), taken during February 2012, and another for the \emph{blue configuration} (BC), taken during April 2012. The detector was constantly operated throughout both periods at a pressure of $10.0$~bar. The TPC cathode was held at $-32$~kV and the gate at $-12$~kV, resulting in a drift field of $667~\mathrm{V\cdot cm^{-1}}$ and an EL field of $2.4~\mathrm{kV\cdot cm^{-1}\cdot bar^{-1}}$. The detector was, in general, operated without interruption with sparks in the EL region no more frequent than one per day. The frequency of these sparks reduced with improved gas quality reducing to a level where distortion of the data was minimal.

During the UVC period, the S1 signals (primary scintillation) were not sufficiently intense to be used as an efficient trigger. Instead, the detector was triggered on the delayed coincidence between a signal in the NaI and an S2 signal (secondary scintillation) in the central PMT of the energy plane up to $400~\mu$s later (a complete drift length plus a margin). Additionally, the NaI signal was required to have a charge greater than 5 photoelectrons, and S2 signal to have an integrated charge greater than $150$ photoelectrons and a duration of between $0.5~\mu$s and $10~\mu$s to ensure that they corresponded to the detection of the 511-keV gammas and not to the abundant high energy signals from atmospheric muons. The trigger rate with this particular set of conditions was 1.5~Hz, consistent with the rate expected from the (weak) calibration source according to the detector simulation. Triggering on the S2 signal demanded that a long ($450~\mu$s) pre-trigger section of the PMT waveforms (buffered by the front-end electronics) be saved to allow the location of the S1 signals (which always precede the S2) during the offline data analysis.

The higher intensity of the S1 signals under the BC and several improvements in the \emph{firmware} of the front-end electronics made it possible to trigger the detector during the second period on the coincidence (within 25~ns) of a NaI signal and an S1 in the energy plane. The integrated charge of the S1 was required to be in the range of 1 to 20 photoelectrons. Under these conditions the trigger rate rose to 2.5~Hz. However, the data size shrank by a factor of 2, since the long pre-trigger was no longer required.

%%%%%%%%%%%%%%%%%%%%%%%%%%%%%%%%%%%%%%%%%%%%%%%%%%%%%%%%%%%%
\subsection{Calibration of the photomultipliers} \label{subsec:PMTCalibration}
\begin{figure}
\centering
\includegraphics[width=0.495\textwidth]{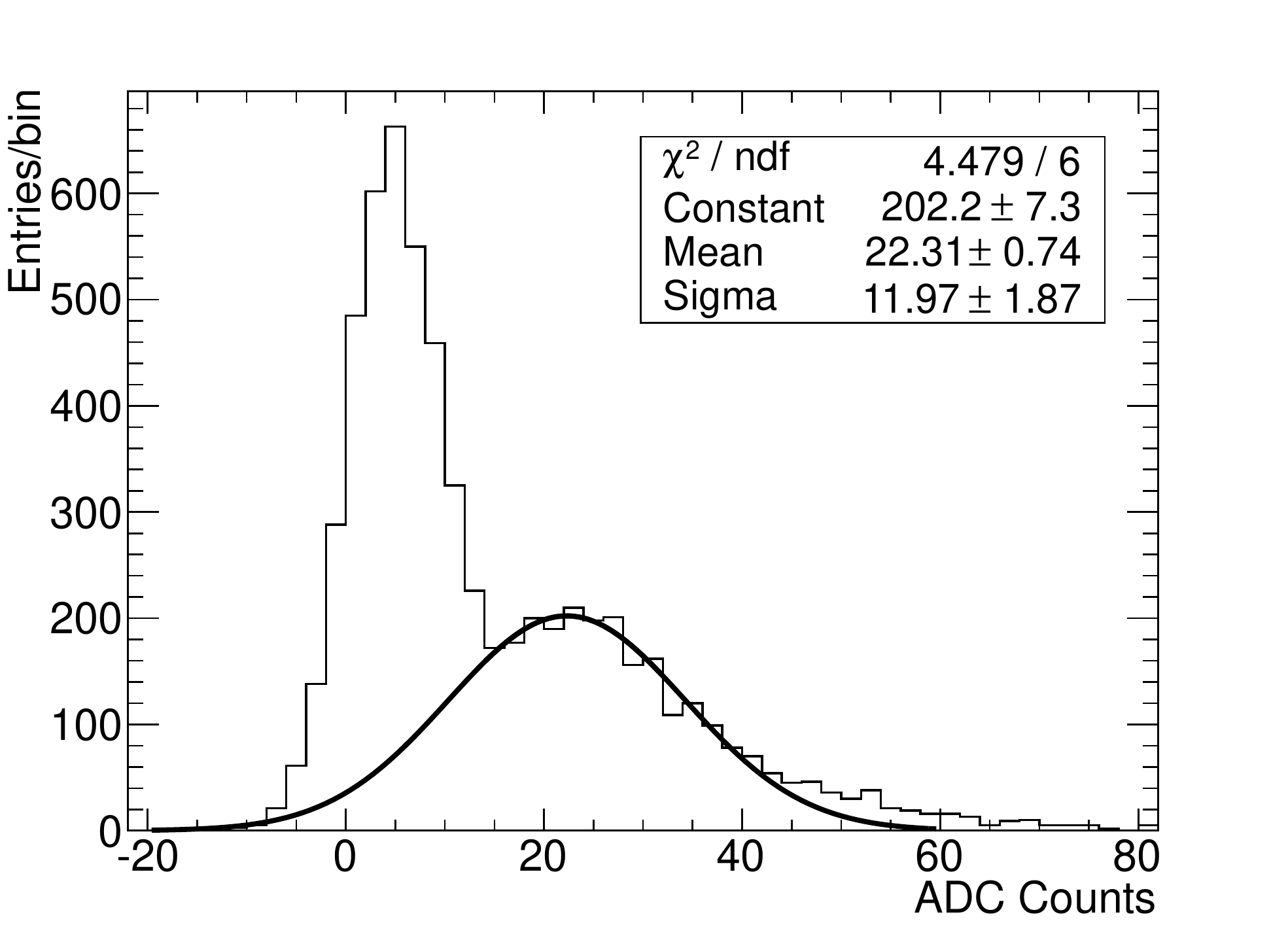}
\includegraphics[width=0.495\textwidth]{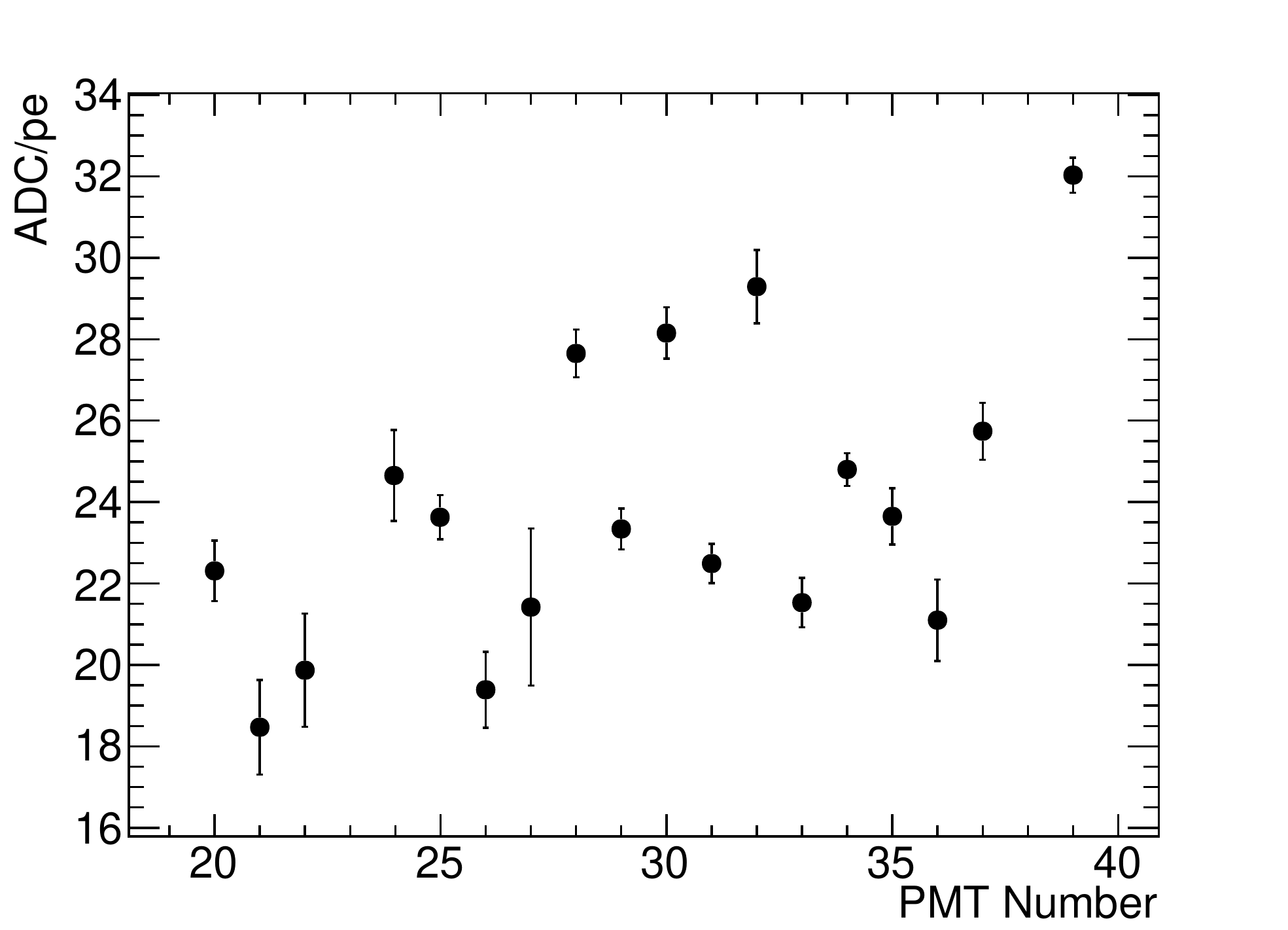}
\caption{Left: Typical single photoelectron spectrum of one of the energy-plane photomultipliers. The single photoelectron peak is clearly separated from the background peak and can be fitted to a gaussian to measure the operating gain and charge resolution of the PMT. The non-gaussian tail on the right side of the peak is due to two-photoelectron events. Right: Gain of the 19 PMTs in the NEXT-DEMO energy plane.
} \label{fig:SpeSpectrum}
\end{figure}

%%%
The 19 photomultiplier tubes in the energy plane were operated at a gain of around $5\times10^6$. The exact gains of the PMTs were measured using a blue LED installed in the tracking plane to illuminate the energy plane with dim light so that only single photoelectron (\emph{spe}) pulses could be recorded. The recorded spe spectrum of each PMT was then fitted to a gaussian to obtain the actual gain, including the conversion factor of the digitization, as illustrated in figure~\ref{fig:SpeSpectrum}. This procedure was repeated regularly during the operation of NEXT-DEMO with no notable changes in the gains of the PMTs.

%%%%%%%%%%
%%%%%%%%%%

The photomultipliers of the tracking plane were operated at a lower gain, $\sim2\times10^6$, to avoid saturation (they receive much more light, being closer to the EL region). At this lower gain, it was not possible to separate the spe peak from the background using the procedure described above. Therefore, the gain of the PMTs at the operating voltage was measured with an oscilloscope in a black box prior to their introduction into the chamber. Their behavior during the run was monitored periodically by comparing the relative positions of the peaks in their individual recorded spectra. This proved to be sufficient since these PMTs were not used to measure the energy of the event.

%%%%%%%%%%%%%%%%%%%%%%%%%%%%%%%%%%%%%%%%%%%%%%%%%%%%%%%%%%%%
\subsection{Waveform processing and event selection} \label{subsec:Waveforms}
\begin{figure}
\centering
\includegraphics[width=0.5\textwidth]{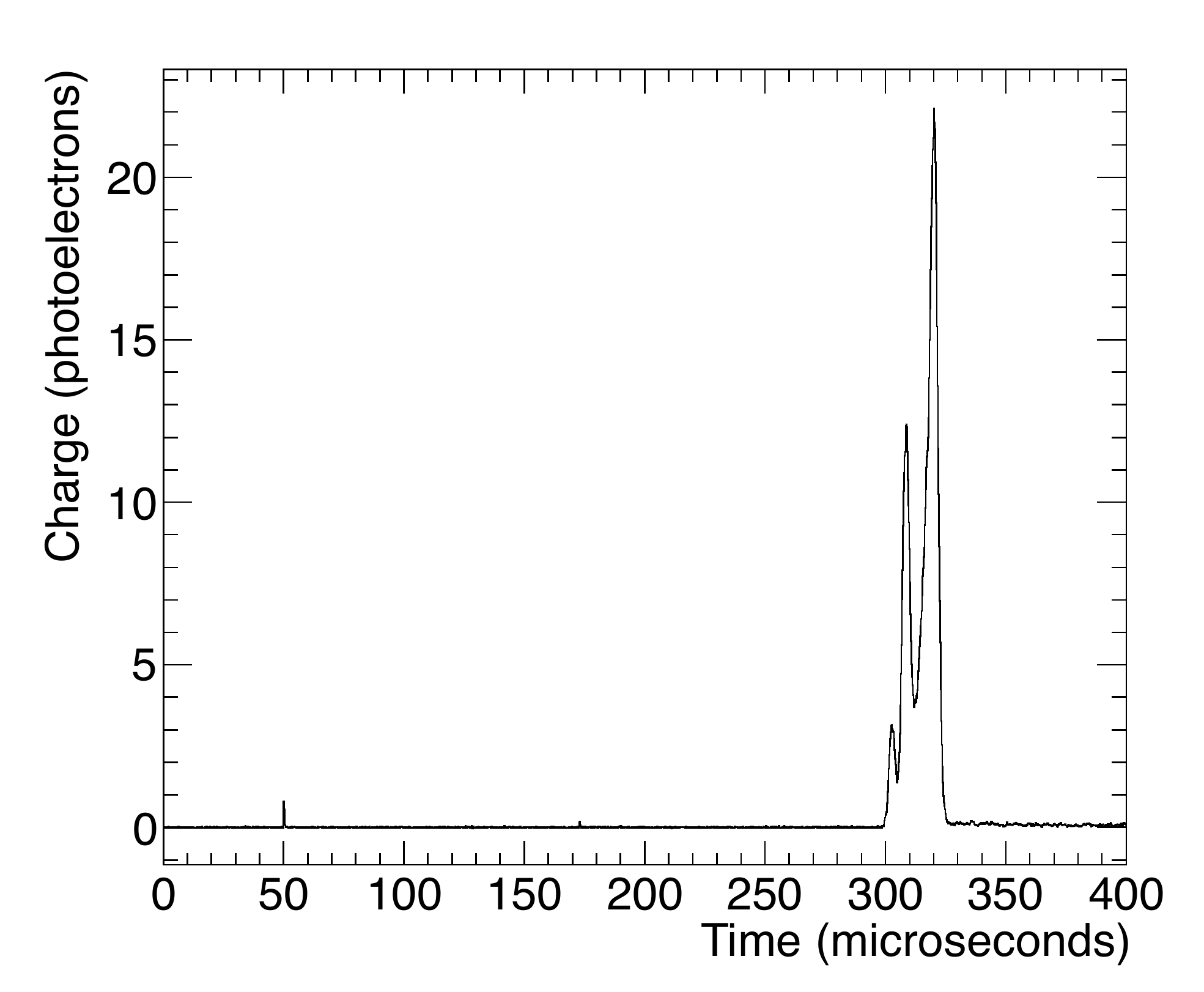}\\
\includegraphics[width=0.45\textwidth]{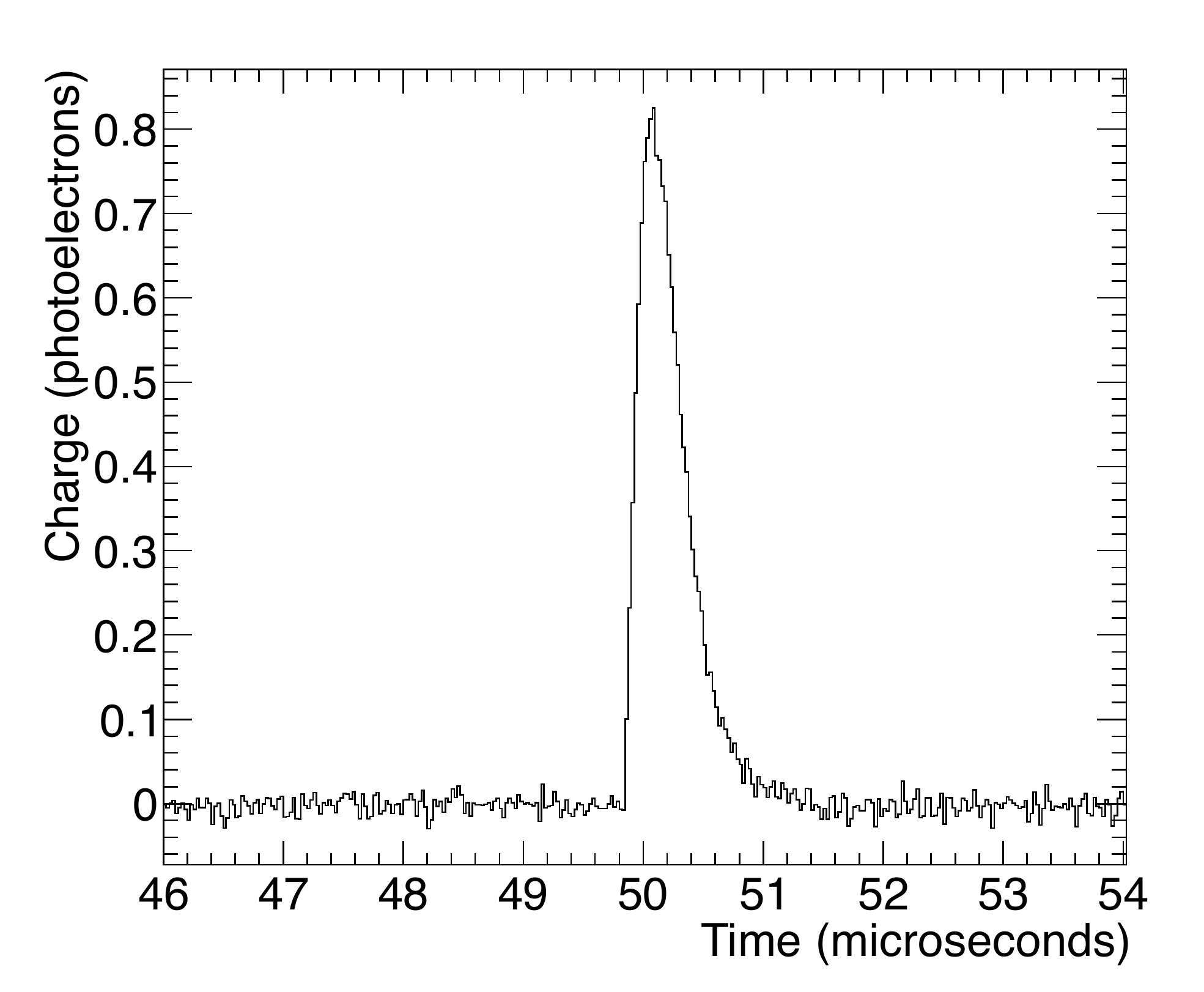}
\includegraphics[width=0.45\textwidth]{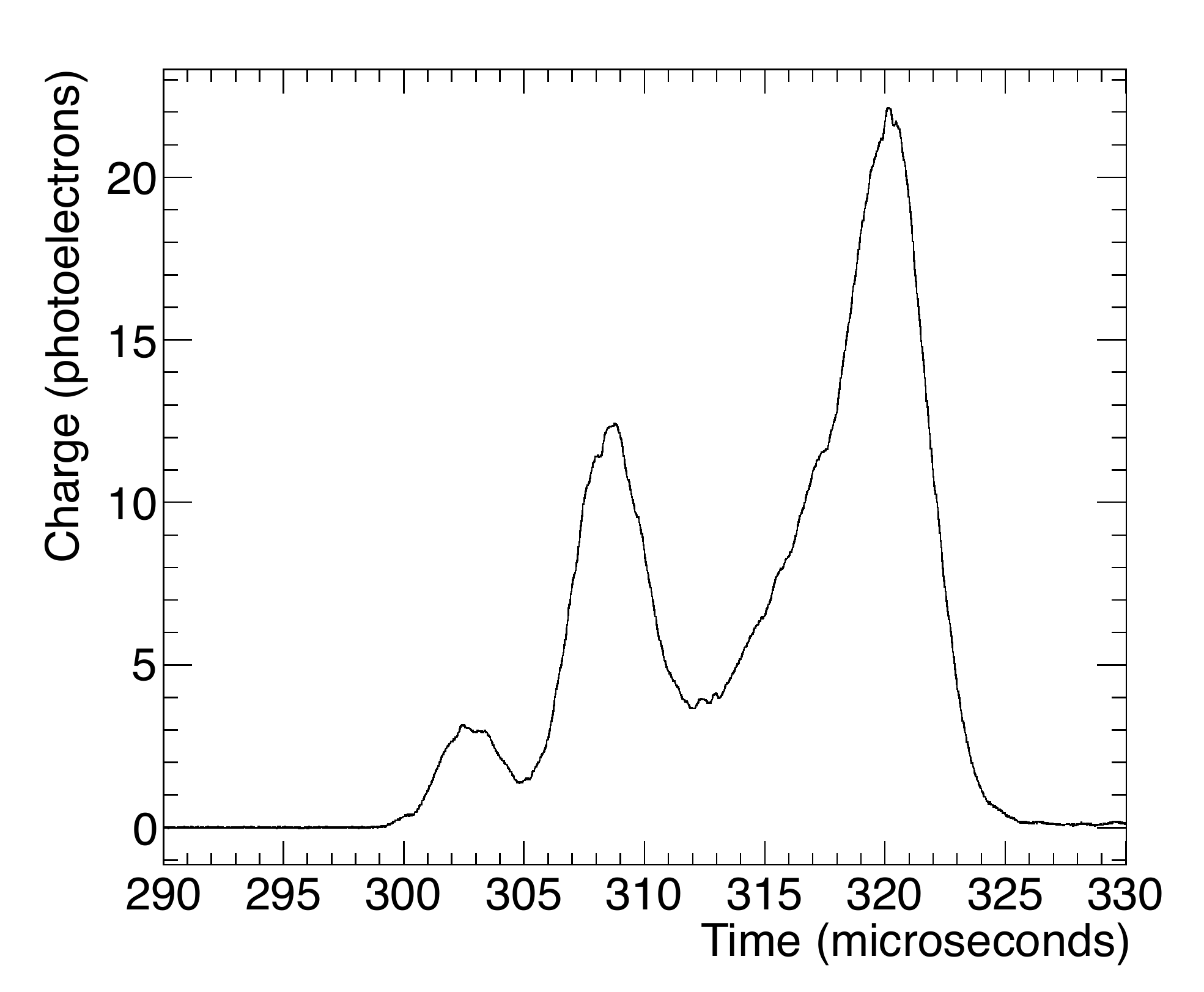}
\caption{Top: Typical waveform of a 511-keV gamma-ray event obtained by summing the pedestal-subtracted, gain-corrected signals of the energy plane (and dividing by the number of PMTs, 19). Bottom: Close-up of the S1 (left) and S2 (right) pulses in the above waveform. The time difference between the leading edges of these pulses indicates the drift time, and hence the longitudinal coordinate of the event. S1 pulses are typically shorter than 3~$\mu$s (this width mostly due to the shaping of the front-end electronics) and have an average charge of at least 0.5~pe/PMT, whereas S2 peaks are longer than 3 $\mu$s and their integrated charge is at least $10~\mathrm{pe/PMT}$.} \label{fig:waveforms}
\end{figure}

%%%
The recorded raw data --- that is, the individual PMT \emph{waveforms} in ADC counts --- were processed in a series of steps. First, ADC pedestals were subtracted for all individual waveforms. The pedestal level was computed for each channel and each event using the first $10\,000$ samples in a waveform that were not considered statistical outliers. Second, using the (channel-dependent) gain calibration constants obtained with the methods described in the previous section, the pedestal-subtracted waveforms were converted from ADC counts to photoelectrons (pes). In order to limit the effect of statistical fluctuations and uncorrelated noise the mean response of the 19 PMT channels re-binned to a 10~MHz sample rate (25\% of the electronics sample rate) is used for event selection and energy reconstruction. Event selection was performed using a peak-finding algorithm which groups collections of consecutive samples rising above the baseline, and classifies these groups as S1-like or S2-like according to their width and total integrated charge. Peaks were considered S1-like (figure~\ref{fig:waveforms}, bottom left) if their width was less than $3~\mu$s and their total charge was at least 0.5~pe/PMT, whereas peaks classified as S2-like (figure~\ref{fig:waveforms}, bottom right) were at least $3~\mu$s wide and had a minimum integrated charge of 10~pe/PMT.

%%%%%%%%%%
%%%%%%%%%%

Events without an S1-like peak or with more than one were discarded at this stage. Events were also rejected if the S1 was not synchronized within a 50~$\mu$s wide window with the NaI signal. For the BC data this check was redundant because the detector had been triggered on that coincidence, but it was kept nonetheless so that both data sets were processed with the same offline algorithms.

%%% SECTION 4. PRIMARY SIGNALS
%%%%%%%%%%%%%%%%%%%%%%%%%%%%%%%%%%%%%%%%%%%%%%%%%%%%%%%%%%%%
\section{Primary signals in NEXT-DEMO} \label{sec:S1S2}
%%%
Charged particles interacting with the xenon gas lose their energy through two different atomic processes: \emph{excitation} and \emph{ionization}. Atomic de-excitations and the recombination of ionization pairs lead to the emission of (primary) scintillation light with a wavelength spectrum that peaks in the vacuum ultraviolet, around 172~nm. Several measurements of the average energy spent in the creation of one primary scintillation photon, $W_\mathrm{sci}$, have been reported for xenon gas: \mbox{$72\pm6$~eV~\cite{Fernandes:2010gg}}, \mbox{$111\pm16$~eV~\cite{DoCarmo:2008}} and $76\pm12$~eV~\cite{Parsons:1990hv}. Their error-weighted mean is \mbox{$W_{\mathrm{sci}}=77\pm5$~eV}. These measurements were made using drift fields of approximately 0.5~kV/cm and gas pressures in the range of 1--20 bar. $W_{\rm sci}$ is correlated with the average energy spent in the creation of an ionization pair, measured as $W_{\rm ion}= 22.0 ~\mathrm{eV}$~\cite{Ahlen:1980xr}. These two empirical values allow the calculation of the expected average ionization and scintillation yields for an electron of energy $E=511$~keV at $N_{\rm sci} = E/W_{\rm sci} = (6640 \pm 430)$ scintillation photons and $N_{\rm ion} = 23\,227$ ionization pairs. Only a small fraction of the scintillation photons is detected in practice due to the finite photocathode coverage of any detector and the inefficiencies in the collection of light. 

The statistical fluctuations of the number of ionization pairs are smaller than those expected from a Poisson distribution, as described by Fano's theory \cite{Fano:1947zz}. Specifically, the variance of $N_{\rm ion}$ can be expressed as
%%%
\begin{equation}
\sigma^{2}_{\rm ion} = F\, N_{\rm ion}\, ,
\end{equation}
%%%
where $F=0.15$ \cite{Nygren:2009zz} is the so-called \emph{Fano factor} of xenon gas. Therefore, the lowest achievable energy resolution for the measurement of the ionization of xenon is
%%%
\begin{equation}
\delta E/E\, = 2.35\, \sqrt{F\, W_{\rm ion}/E}\,.
\end{equation}
%%%

Under the influence of the electric field applied to the detector, ionization electrons drift towards the anode at a velocity of approximately 1~mm$/\mu$s~\cite{Alvarez:2012hu}, while positive ions drift towards the cathode. Electrons are absorbed by electronegative impurities in the gas at a rate described by:
%%%
\begin{equation}
N(t_{\rm D}) = N_0\cdot\exp(-t_{\rm D}/\tau)\, , \label{eq:ExpAttachment}
\end{equation}
%%%
where $t_{\rm D}$ is the drift time and $\tau$ is the mean \emph{electron lifetime} in the gas. During drift the electons also diffuse in 3 dimensions which results in a physical limit on the position resolution dependent on the drift field.

Once the primary electrons reach the EL region, they are accelerated between two parallel meshes giving rise to secondary scintillation (or electroluminescence). The number of photons emitted per primary electron and per unit of drift length and pressure, the so-called \emph{reduced electroluminescence yield}, is linearly proportional to the reduced electric field $E/p$ above a threshold of about $0.83~\mathrm{kV~cm^{-1}~bar^{-1}}$ and up to $\sim6~\mathrm{kV~cm^{-1}~bar^{-1}}$ \cite{Monteiro:2007vz}:
%%%
\begin{equation}
\frac{Y}{p} = 140\ \frac{E}{p} - 116\ \left[\mathrm{photons~electron^{-1}~cm^{-1}~bar^{-1}}\right] \label{eq:yopfit}
\end{equation}
%%%
The fluctuations associated with the electroluminescence production are much smaller than the Fano factor and, hence, the energy resolution of the detector is proportional to the fluctuations in ionization and in the response of the photodetection system~\cite{Nygren:2009zz, Oliveira:2011xk}.

In NEXT-DEMO the width of the EL region (delimited by the gate and anode meshes) is 0.5~cm, and for both the UVC and BC periods the operating pressure was 10 bar and the voltage across the grids was 12~kV. These parameters correspond to a reduced electric field of $2.4~\mathrm{kV~cm^{-1}~bar^{-1}}$, or equivalently, according to equation~(\ref{eq:yopfit}), 1\,100 photons per primary electron emitted across the EL region of NEXT-DEMO. 

%%%%%%%%%%
\begin{figure}
\centering
\includegraphics[width=0.495\textwidth]{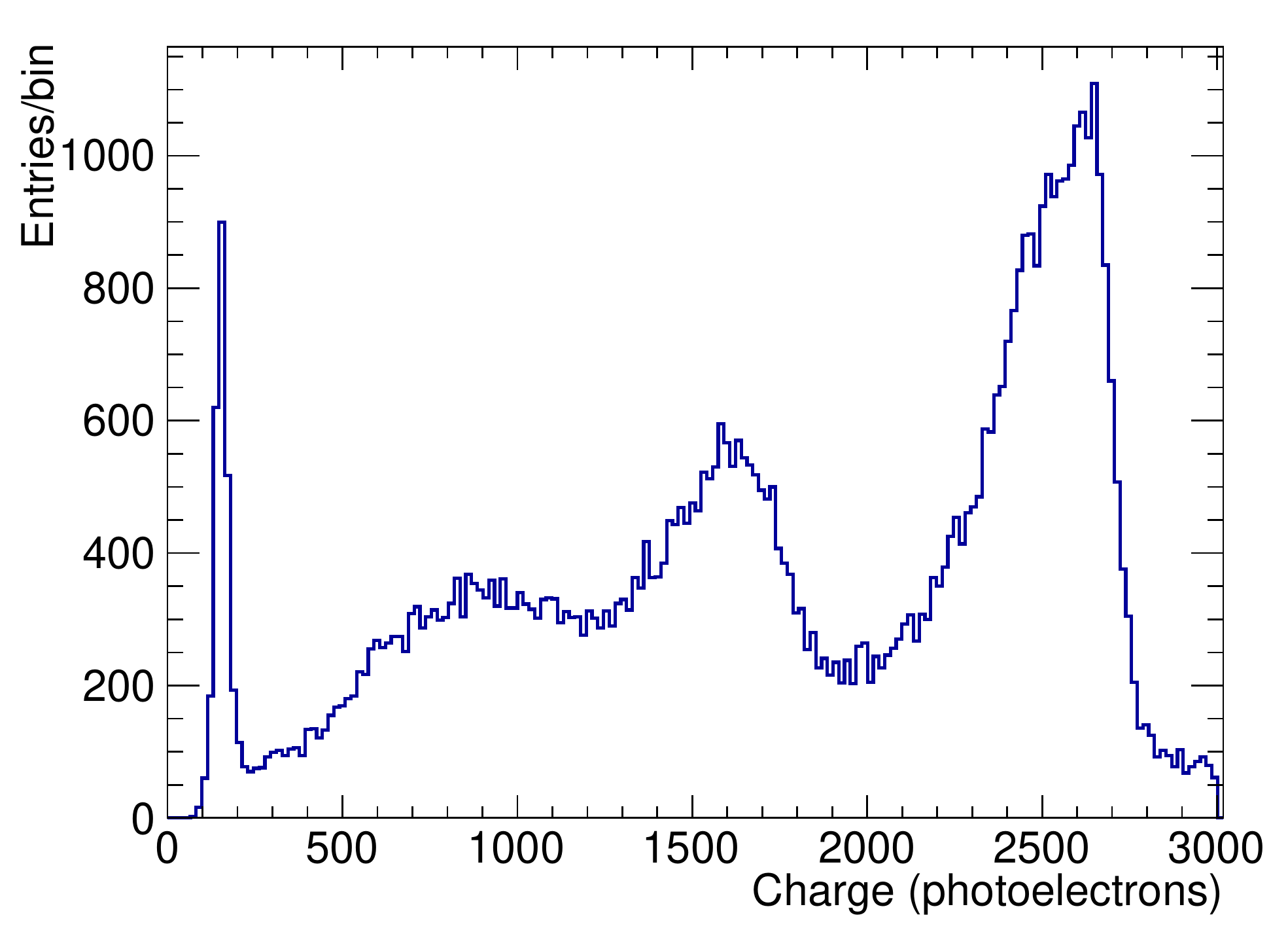}
\includegraphics[width=0.495\textwidth]{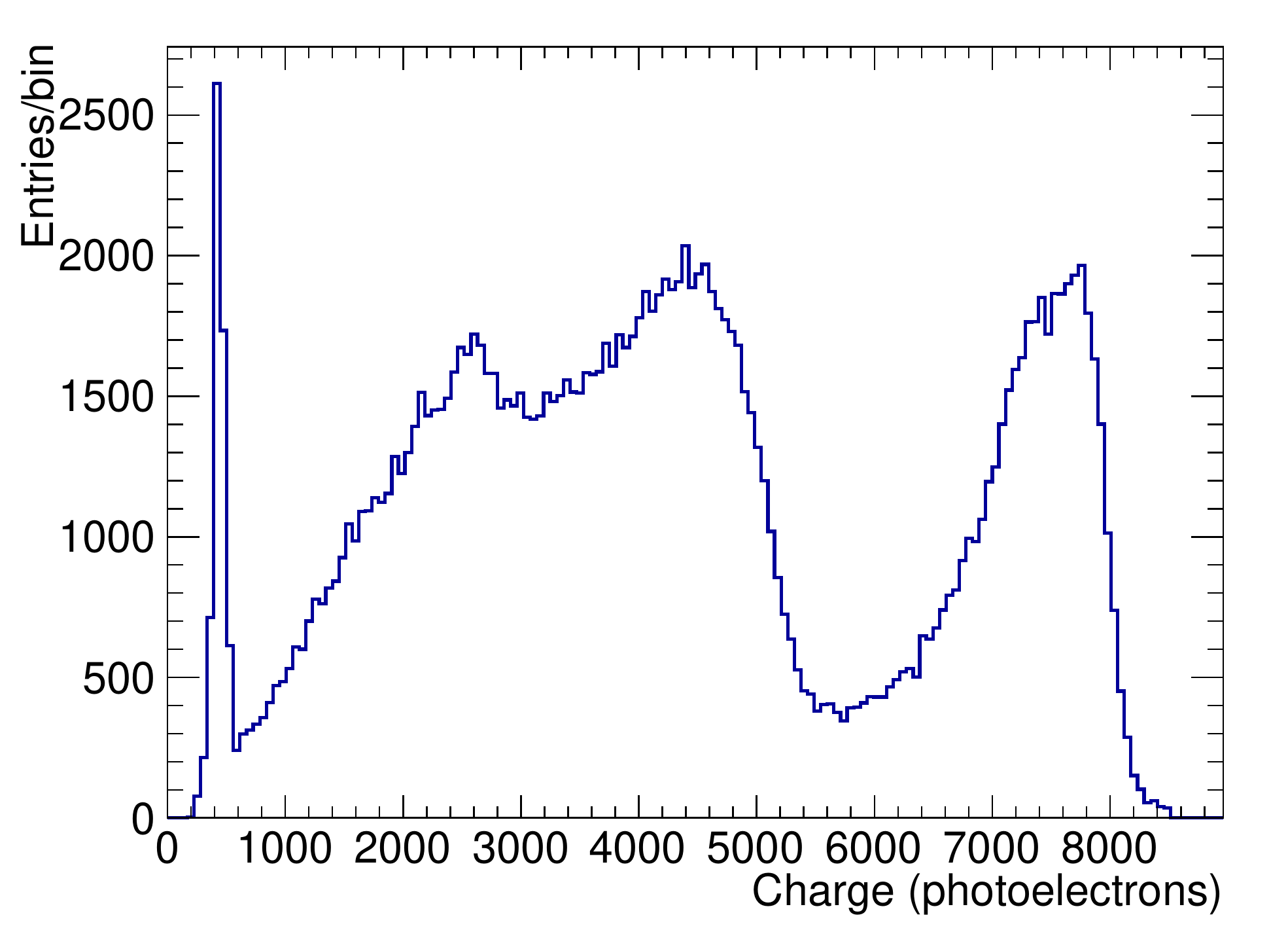}
\caption{Raw energy spectrum --- defined as the total integrated secondary-scintillation (S2) charge measured by the energy plane divided by the number of PMTs --- recorded in NEXT-DEMO for the \emph{ultraviolet} (left panel) and \emph{blue} (right panel) configurations. From left to right in the spectra, the xenon x-ray peak, the Compton edge and the full energy peak are clearly visible in spite of the mediocre energy resolution before corrections. The discrepancy between the two spectra in the relative size of the Compton edge and the photoelectric peak is due to the different configuration of the detector trigger for the two periods.} \label{fig:SpectrumRaw}
\end{figure}
%%%%%%%%%%

%%%%%%%%%%
\begin{figure}
\centering
\includegraphics[width=0.495\textwidth]{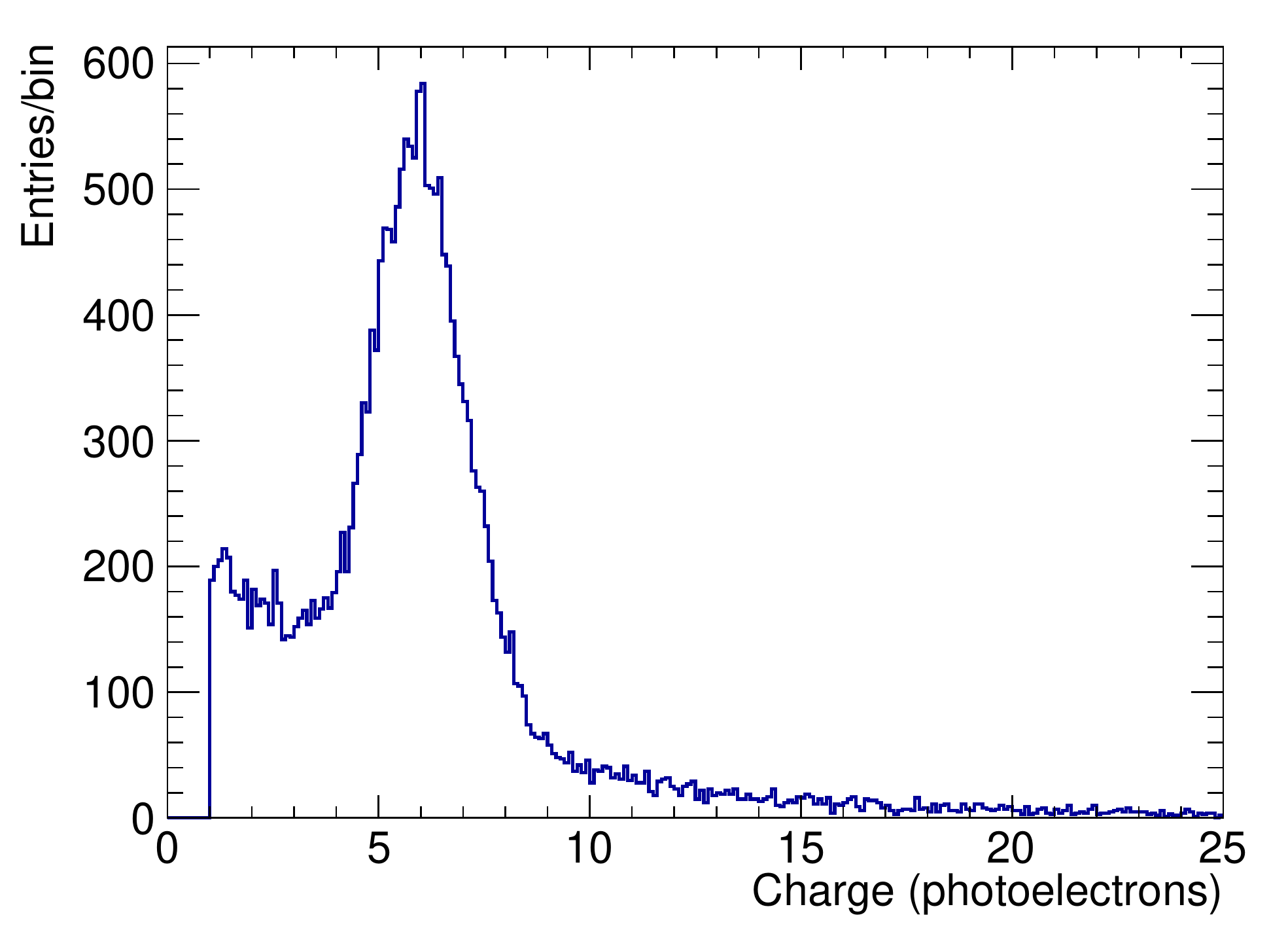}
\includegraphics[width=0.495\textwidth]{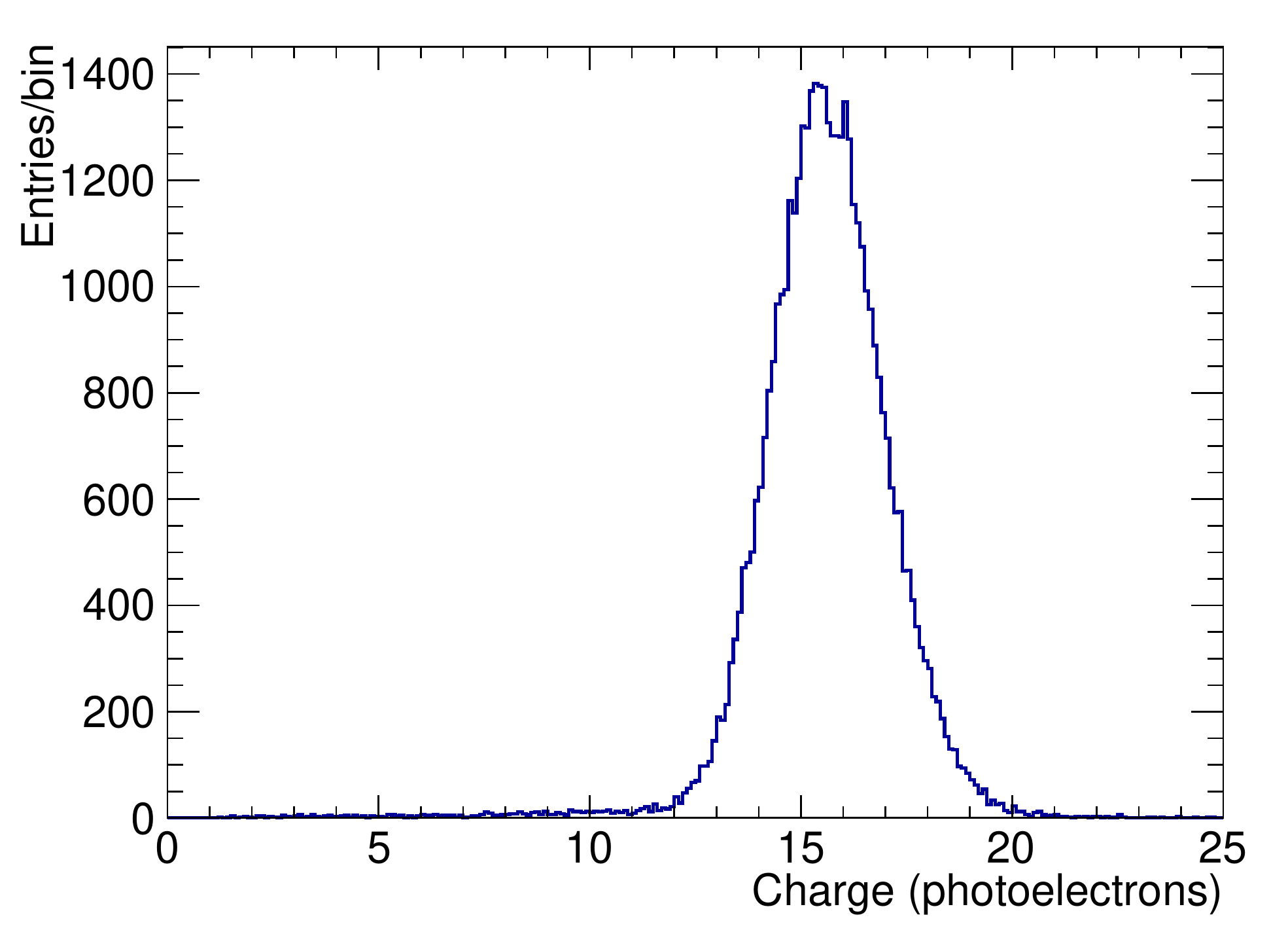}
\caption{Intensity of the primary scintillation (S1) signal --- i.e., integrated charge of the S1 peak of the energy-plane summed waveform --- per PMT for photoelectric (full-energy) events. Left, for the \emph{ultraviolet} configuration; right, for the \emph{blue} configuration.} \label{fig:S1Integral}
\end{figure}
%%%%%%%%%%

As mentioned in section~\ref{sec:Introduction}, the energy plane of NEXT-DEMO detects a fraction of the secondary scintillation to provide a precise measurement of the total energy deposited in the chamber. More precisely, the energy of an event is defined as the total integrated charge of all the S2 peaks normalized to the number of PMTs in the energy plane. Figure~\ref{fig:SpectrumRaw} shows the raw energy spectrum recorded in the chamber for both configurations. The xenon x-ray peak at low energy is clearly visible, as well as the Compton spectrum and the full-energy photoelectric peak. The TPB-coated light tube improves the light collection by a factor of 3 both at the photopeak and for the S1 signal (see~\ref{fig:S1Integral}). This improvement was attributed to the increased quantum efficiency at the emission peak of the TPB  ($\sim430$~nm) and the near-perfect reflectivity of PTFE at this wavelength.

%%%%%%%%%%
\begin{figure}
\centering
\includegraphics[width=0.495\textwidth]{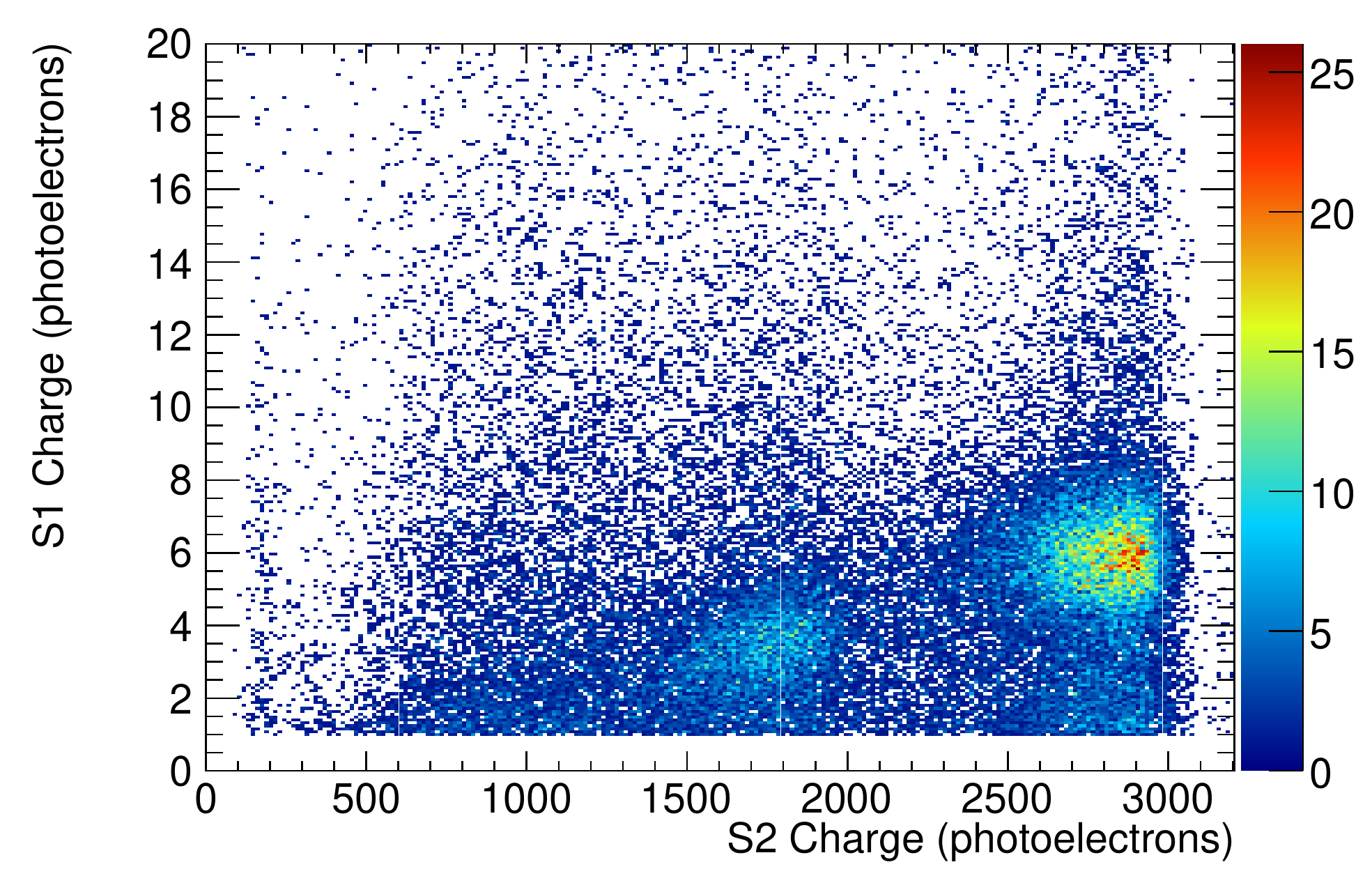}
\includegraphics[width=0.495\textwidth]{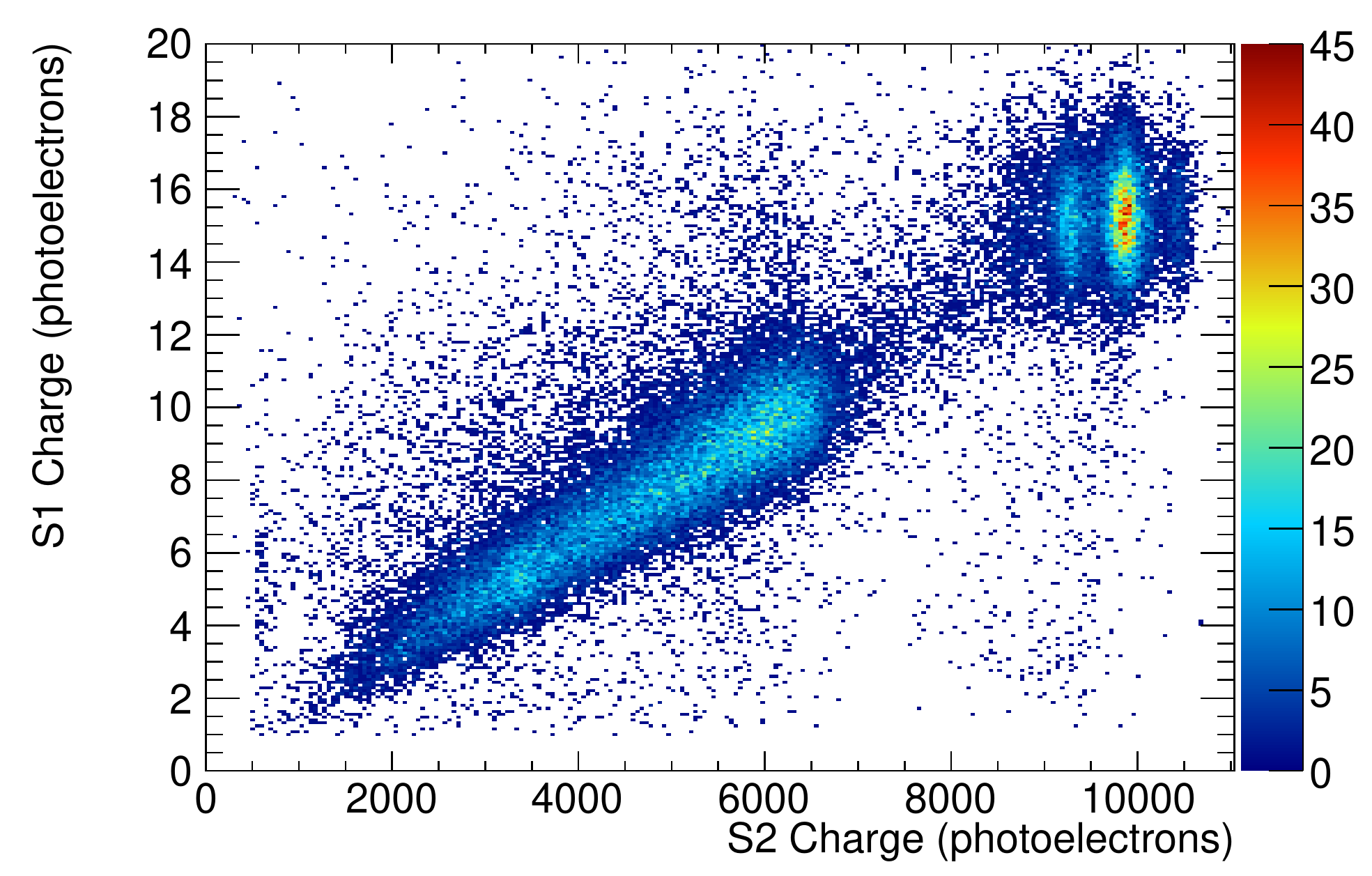}
\caption{Correlation between the total integrated charge (per PMT) of the S1 and the S2 signals (after spatial corrections) for the full spectrum of interactions resultant from the \NA\ source. Left, for the \emph{ultraviolet} configuration; right, for the \emph{blue} configuration. The island of full-energy events shows no indication of correlated fluctuations between S1 and S2.} \label{fig:S1vsS2}
\end{figure}
%%%%%%%%%%

Figure \ref{fig:S1vsS2} shows the correlation between the S1 and S2 signals (after spatial corrections; see section~\ref{sec:EnergyResolution}) for both the UVC and BC. Notice that the island of full-energy (photoelectric) events shows no indication of correlated fluctuations between S1 and S2, like those observed with alpha particles \cite{Alvarez:2012hu} or in liquid xenon \cite{Conti:2003av}.

%%% SECTION 5. ENERGY RESOLUTION
%%%%%%%%%%%%%%%%%%%%%%%%%%%%%%%%%%%%%%%%%%%%%%%%%%%%%%%%%%%%
\section{Energy resolution analysis} \label{sec:EnergyResolution}
%%%
The data of the raw energy spectrum shown in figure~\ref{fig:SpectrumRaw} are affected by several factors that contribute to degrade the intrinsic energy resolution of gaseous xenon. An important factor is the position-dependent response of the detector. In order to compensate for this effect the energy recorded is corrected for according to the reconstructed position in the detector. The reconstructed energy, $E_\mathrm{rec}$, can be parameterized as the product of the measured energy $E_{\rm obs}$ and a certain \emph{detector response function} $\eta$:
%%%
\begin{equation}
E_\mathrm{rec} = E_\mathrm{obs}(x,y,z) \cdot \eta(x,y,z).
\end{equation}
%%%

However, the position of an event is always reconstructed with some finite resolution. The longitudinal position $z'$ can be reconstructed with high precision using the time between S1 and S2 signals and the \emph{drift velocity}, measured to be $\sim 1~\mathrm{mm/\mu s}$ \cite{Alvarez:2012hu}. The $x'$ and $y'$ coordinates are reconstructed using the \emph{barycentre} of the tracking-plane PMTs with recorded signal:
%%%
\begin{equation}
x' = \frac{\sum_i\ A_{i}\ x_{i}}{\sum_{i}\, A_{i}}\, , \quad y' = \frac{\sum_i\ A_{i}\ y_{i}}{\sum_{i}\, A_{i}}\quad {\rm and} \quad R' = \sqrt{x'^2 + y'^2}\ ,
\label{eq:BarX}
\end{equation}
%%%
where $A_i$ is the integrated charge of the S2 measured by PMT $i$, which is positioned at $x_{i}y_{i}$. Since the events considered in this analysis have, in general, an extent in the detector of only a few mm --- the average effective extent, considering multiple scattering, of an electron of 511 keV being about 20~mm --- they are considered point-like for the purposes of energy correction. More energetic (and thus longer) tracks would have to be segmented and a different correction applied to the resulting sections (see section~\ref{sec:Topology} for a preliminary discussion).

%%%%%%%%%%%
\begin{figure}
\centering
\includegraphics[scale=0.35]{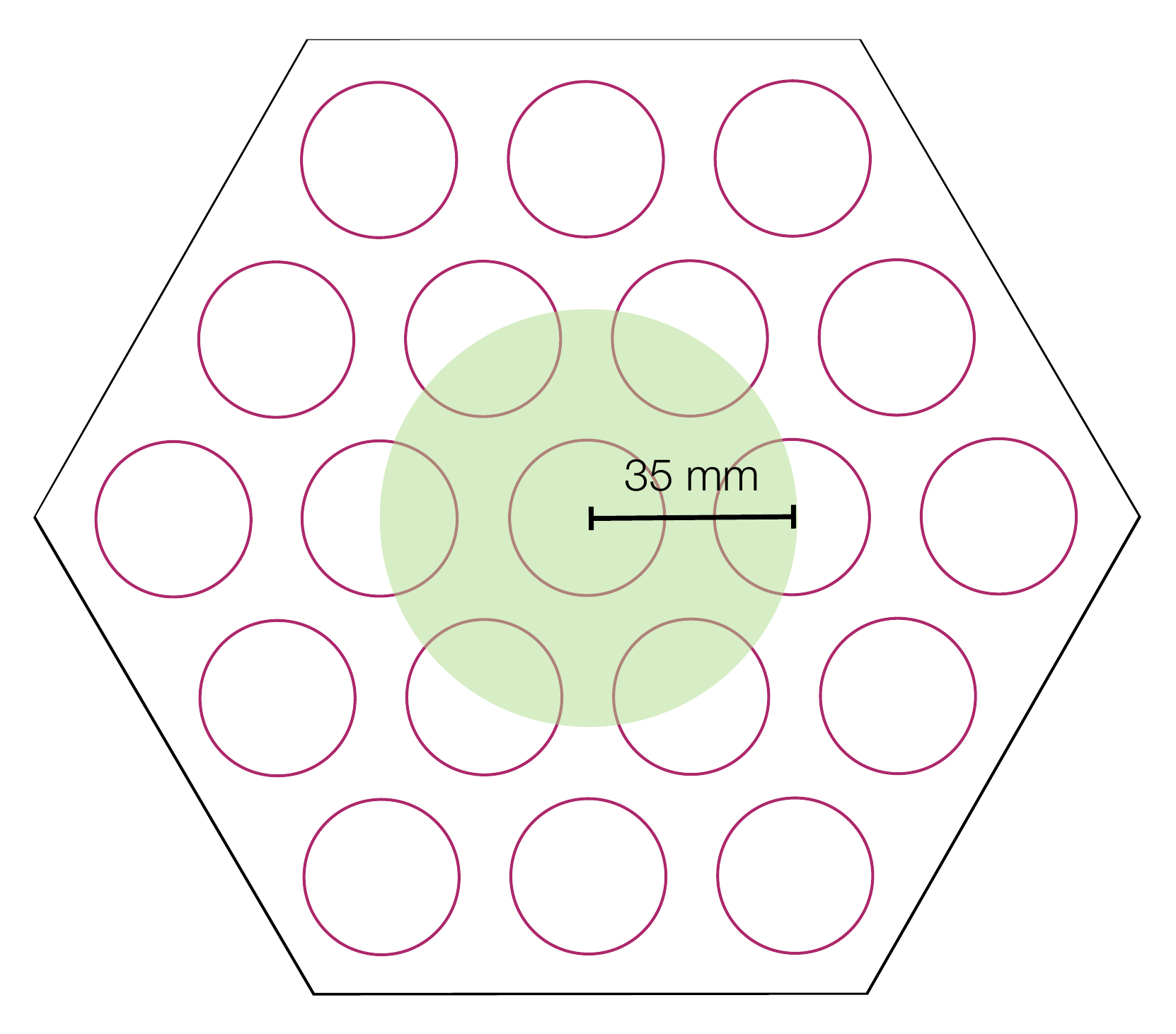}\\
\includegraphics[width=0.495\textwidth]{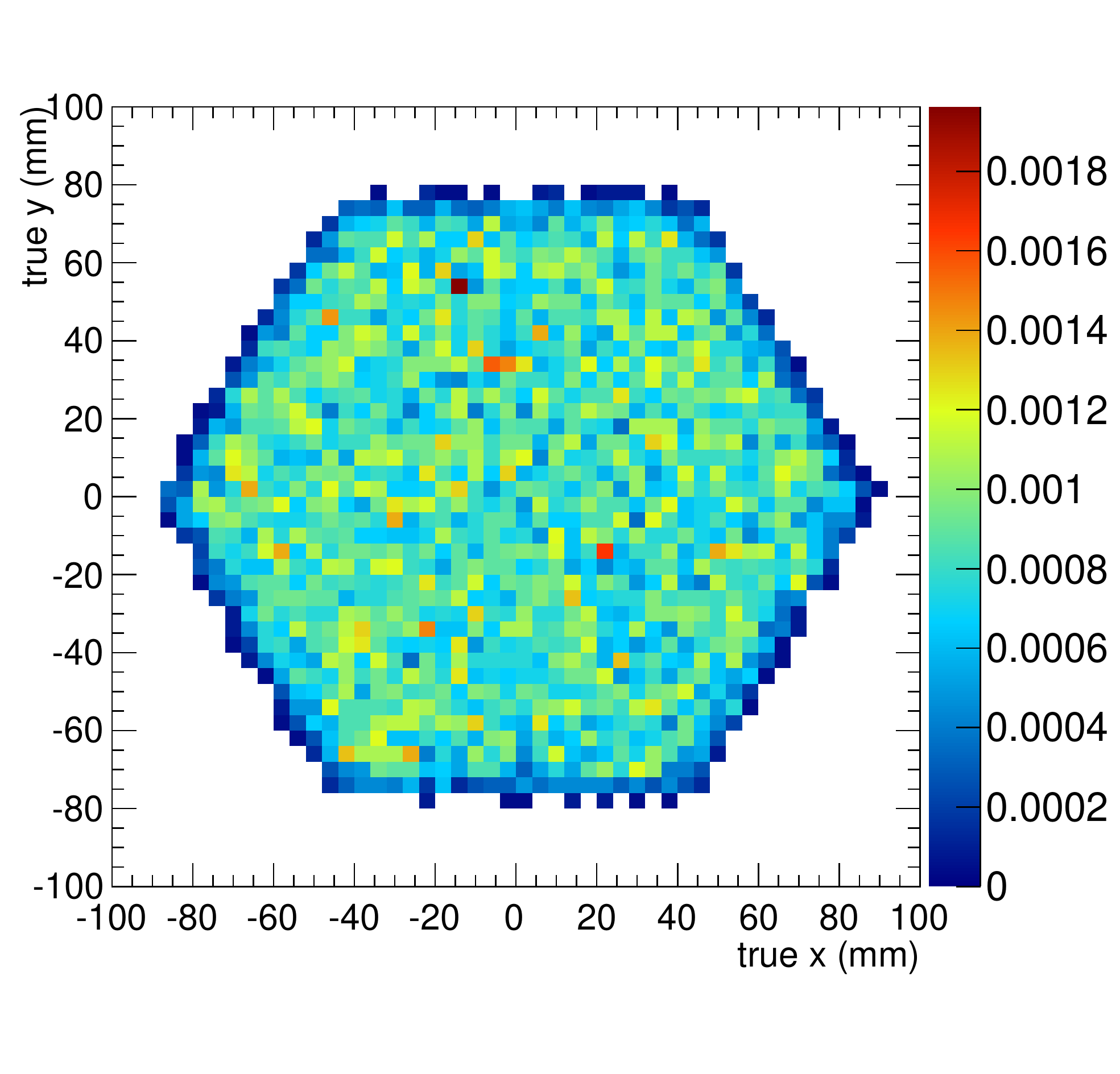}
\includegraphics[width=0.495\textwidth]{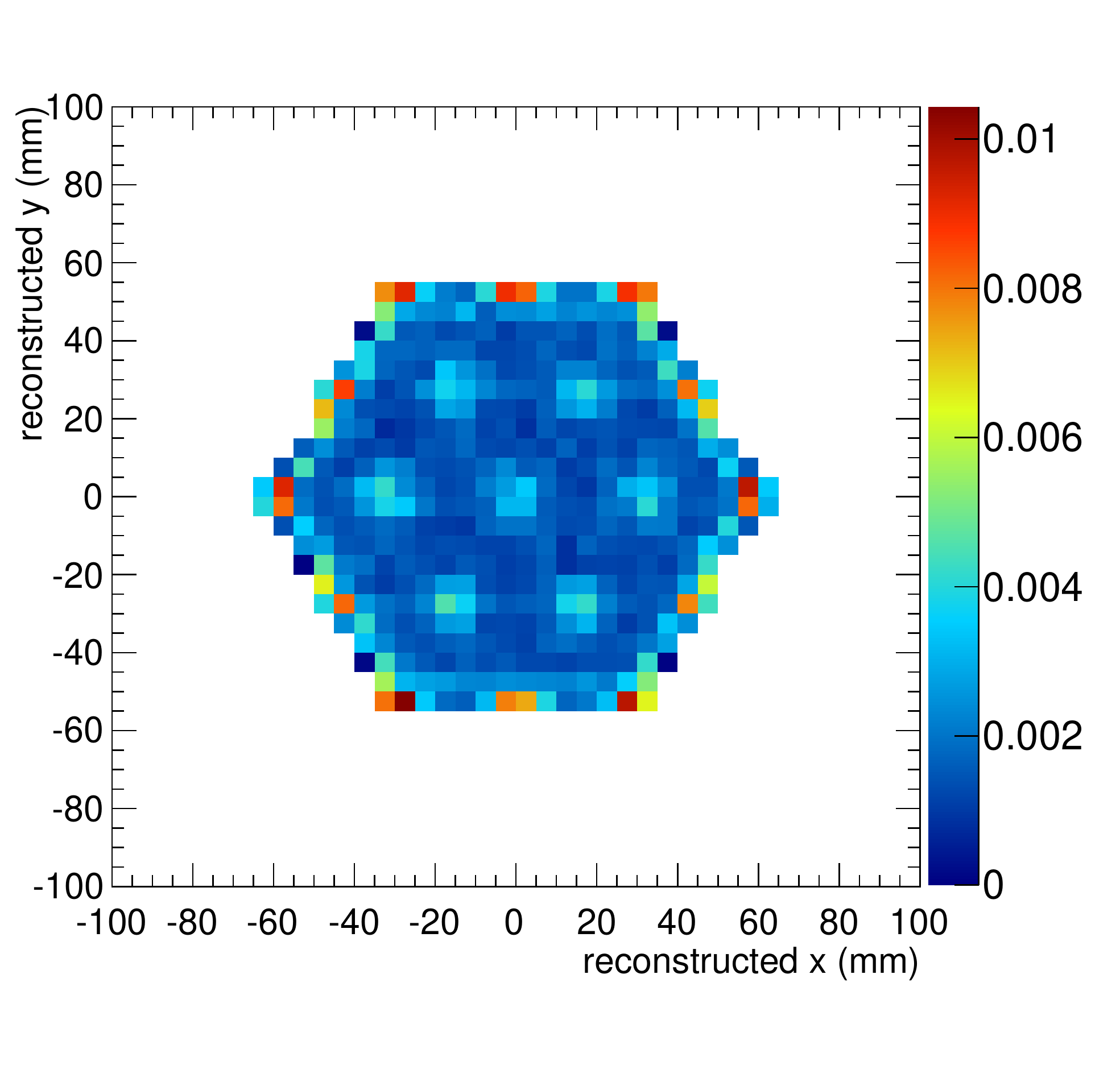}
\caption{Top: Sketch of the NEXT-DEMO tracking plane showing the position of the photomultipliers (magenta circumferences) and the defined fiducial volume (green circle). Bottom: True average position (left panel; see text for definition) of simulated events compared to the reconstructed position (right panel) using the barycenter of the tracking plane. The original uniform distribution of events gets distorted, with events accumulating at the centre of the PMTs and in a region intermediate between the first and second rings of sensors.}  \label{fig:Fiducial} 
\end{figure}
%%%%%%%%%%

The reduced position resolution afforded by the limited number of PMTs in the tracking plane and their large surface area means that the barycentre tends to be reconstructed with a bias towards the centres of the PMTs and detector, as seen in figure~\ref{fig:Fiducial}. For this reason, the energy correction is calculated using events reconstructed in the volume defined by the cylinder centred on the central PMT with cross-section radius $R'<35$~mm (approximately 13\% of the hexagonal cross section instrumented region). In this way the bias has a limited effect on the reconstruction.

The detector response function can be expressed as the product of three terms:
%%%
\begin{equation}
\eta(x,y,z) = Z(z') \cdot \rho(R') \cdot \Phi(\phi')\, , \label{eq:SpatialCorr}
\end{equation}
%%%
where $Z(z')$~is a function that depends only on the reconstructed longitudinal coordinate, $\rho(R')$ is a function that depends only on the reconstructed radius
$R'$, and $\Phi(\phi')$~is a function that depends only on the reconstructed azimuthal coordinate $\phi' = \tan^{-1}({y'/x'})$ with possible factorization of these terms dependent on the amount of correlation between corrections. 

%%%%%%%%%%
\begin{figure}
\centering
\includegraphics[width=0.495\textwidth]{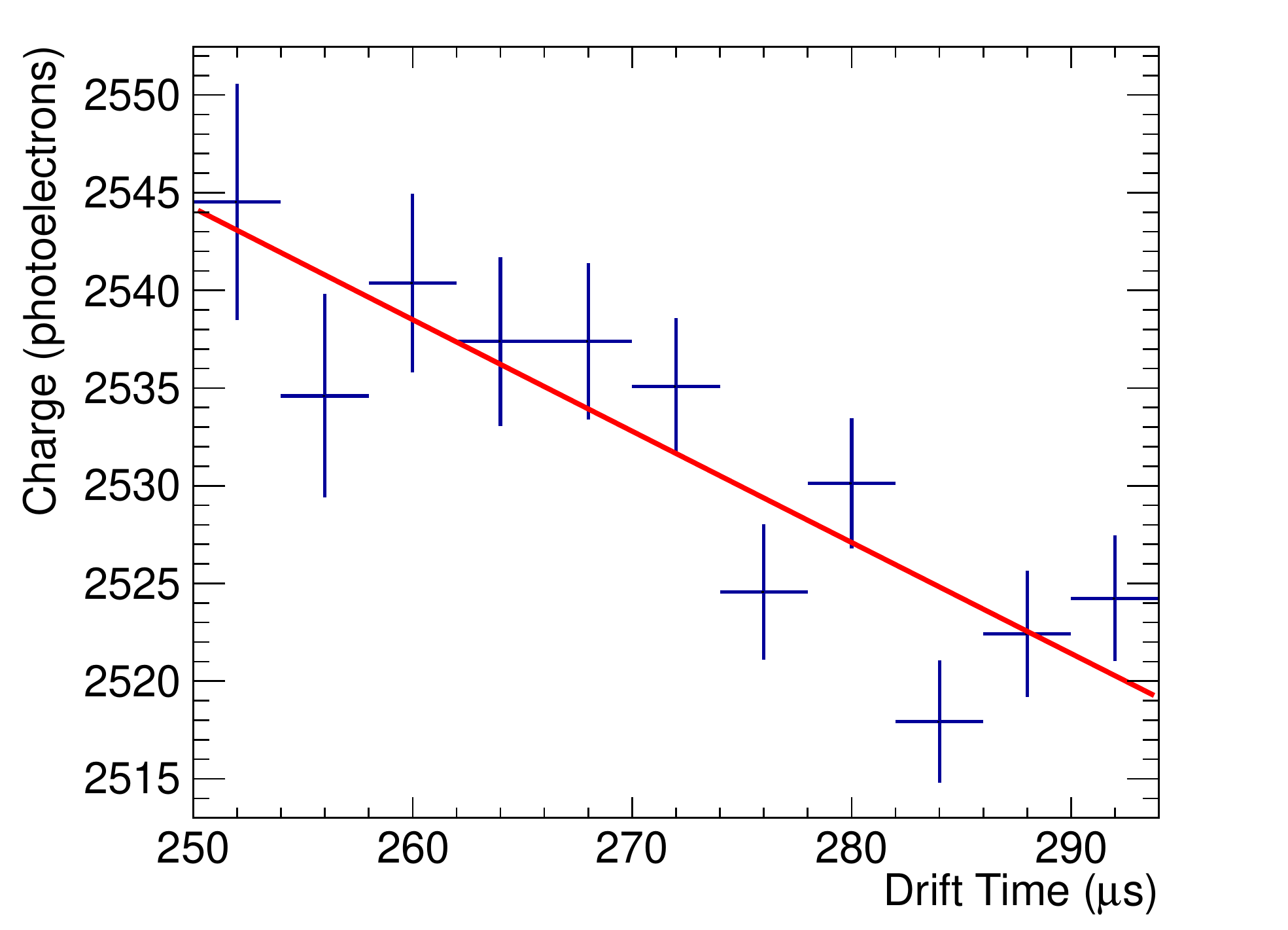}
\includegraphics[width=0.495\textwidth]{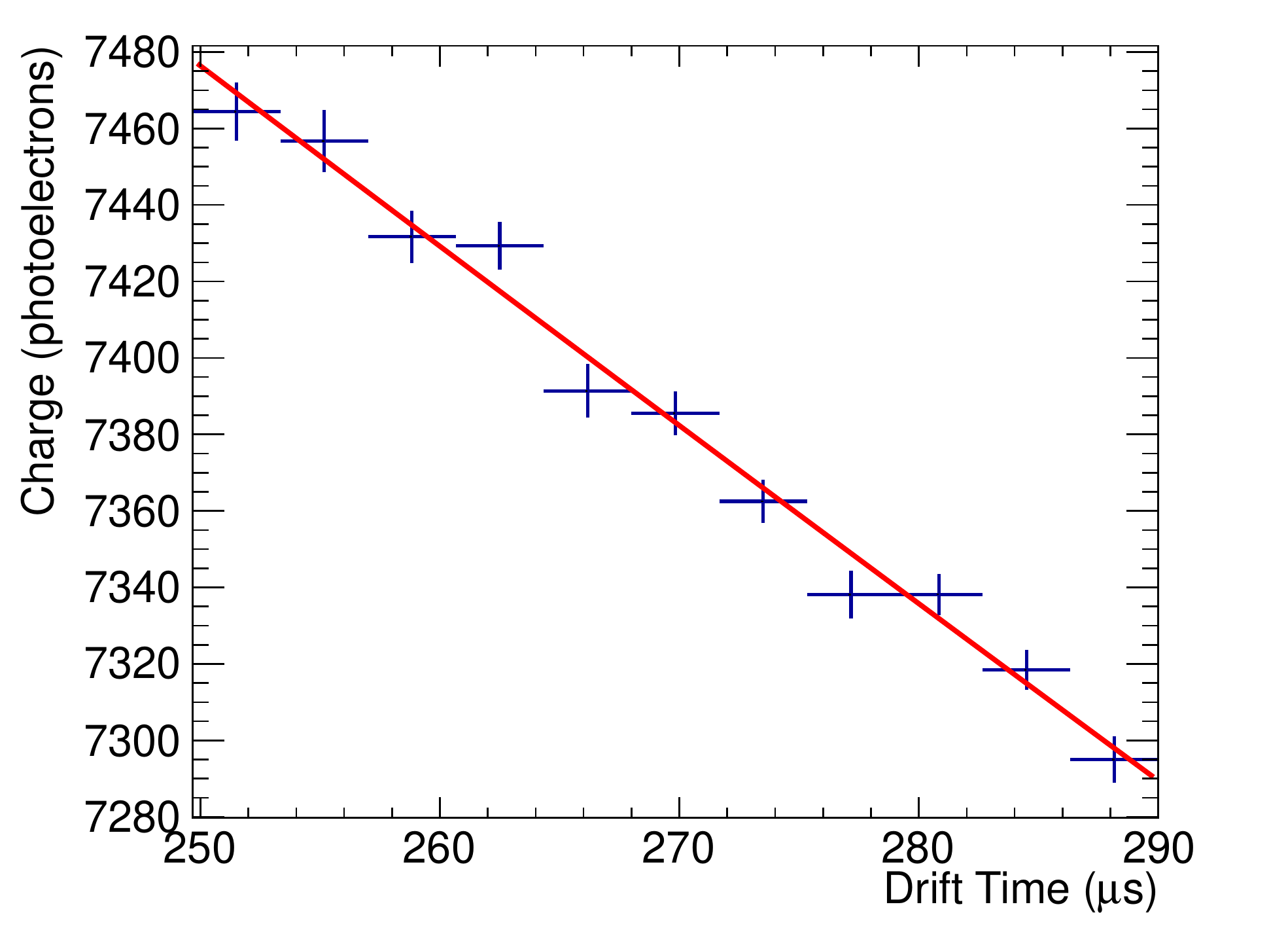}
\caption{Dependence of the total integrated charge (energy) on drift time for photoelectric events. Left, for the \emph{ultraviolet} configuration (UVC); right, for the \emph{blue} configuration (BC). An exponential fit (red line) to the data determines the electron lifetime in the gas, obtaining 4.0~ms for the UVC and 1.7~ms for the BC.} \label{fig:S2vzZ}
\end{figure}
%%%%%%%%%%

%%%%%%%%%%
\begin{figure}
\centering
\includegraphics[width=0.495\textwidth]{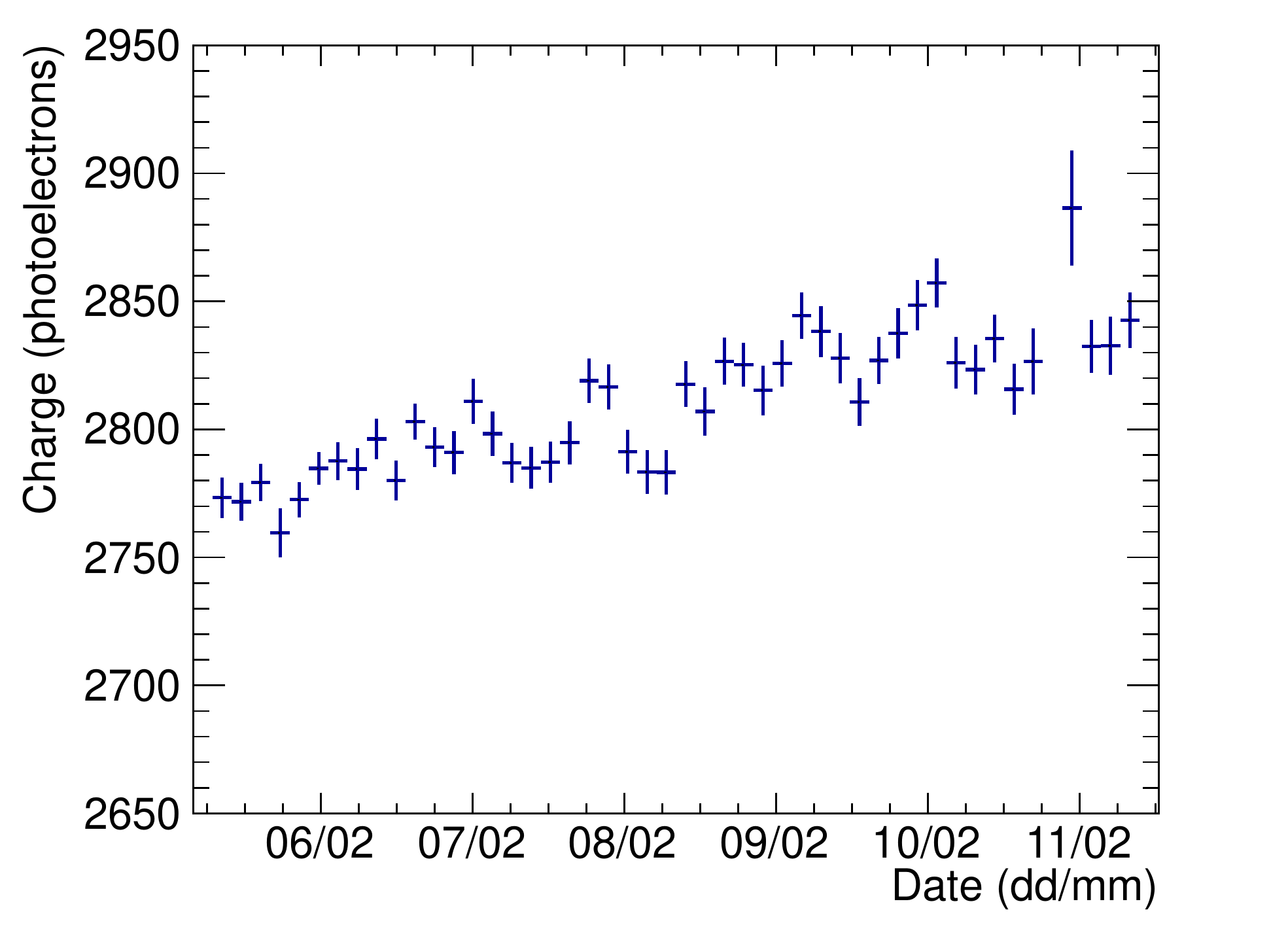}
\includegraphics[width=0.495\textwidth]{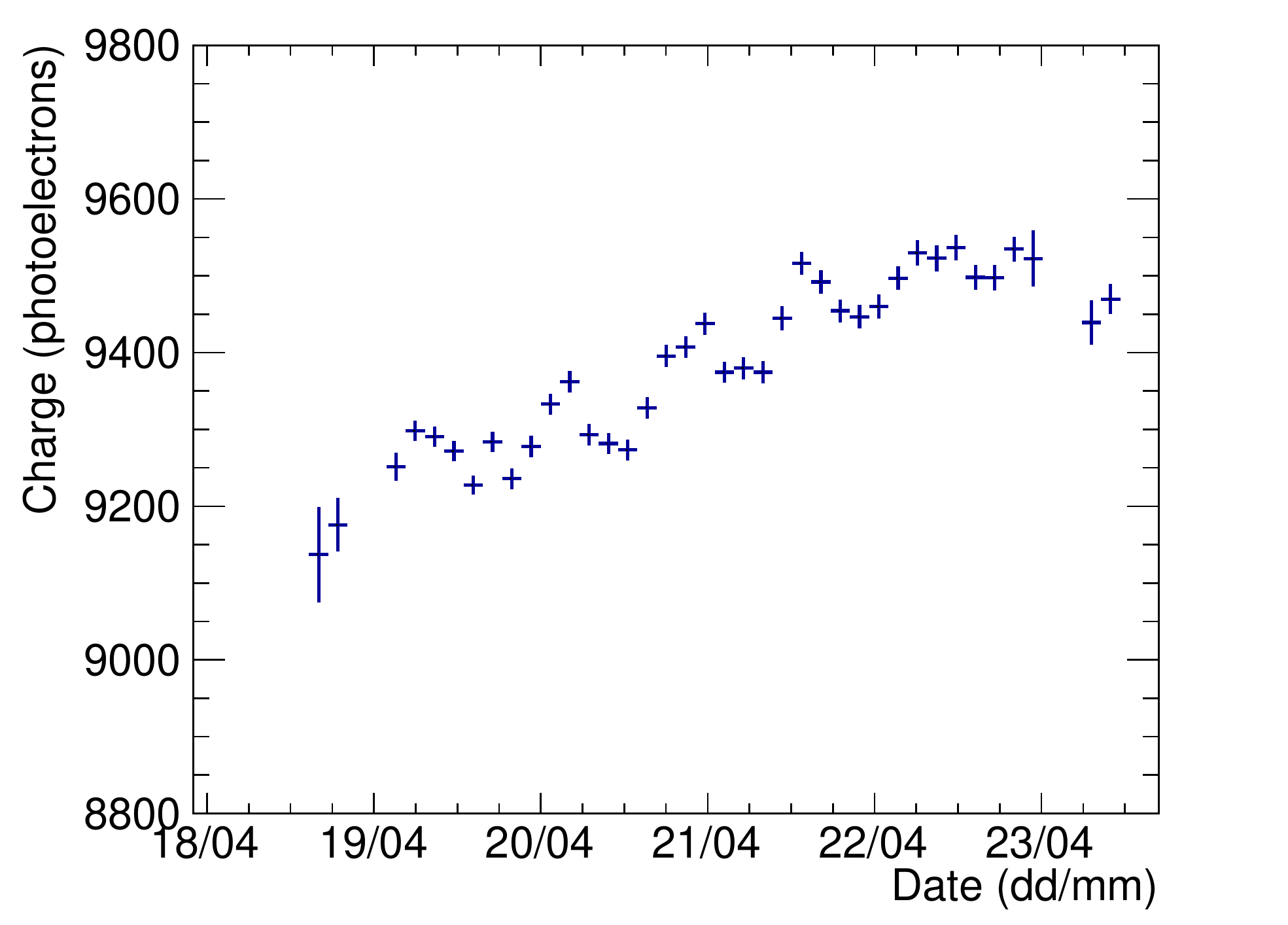}
\caption{Evolution of the average S2 charge of the photoelectric peak during the data-taking period for both the UVC (left) and the BC (right).} \label{fig:EvolutionS2}
\end{figure}
%%%%%%%%%%

Figure~\ref{fig:S2vzZ} shows the dependence of the total integrated charge on the longitudinal coordinate (expressed, in this case, as drift time) for full-energy (photoelectric) events in the interval $[250, 290]~\mu{\rm s}$. This cut was chosen to limit the influence of edge effects at the cathode and the limited statistics that were available for events with drift time less than $250~\mu$s (due to the source port position). Fitting a negative exponential to a set of data allows the determination of the electron lifetime in the gas as demonstrated by equation~\ref{eq:ExpAttachment}. Knowledge of this term enables the calculation of the correction term $Z(z')$ since it is dominated by electron attachment. The data yield $\tau$: $4.0 \pm 0.4$~ms for the UVC and $1.7 \pm 0.4$~ms for the BC. As mentioned in section~\ref{sec:Detector}, new xenon (not clean) had to be added periodically to the detector during the BC runs due to a micro-leak, and hence the quality of the gas was generally worse than that in the UVC, and a greater amount of attachment was expected.

%%%%%%%%%%
\begin{figure}
\centering
\includegraphics[width=0.495\textwidth]{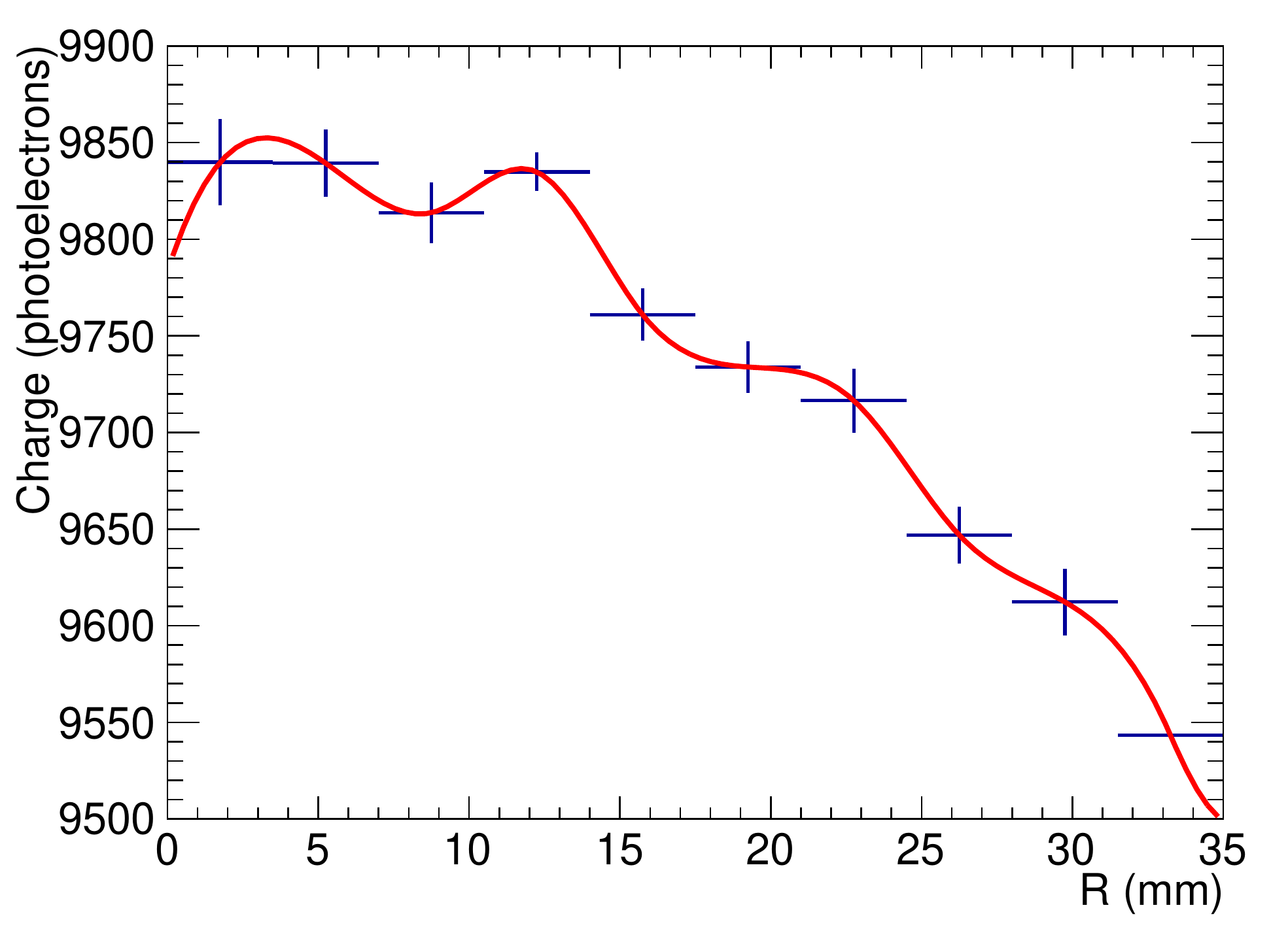}
\includegraphics[width=0.495\textwidth]{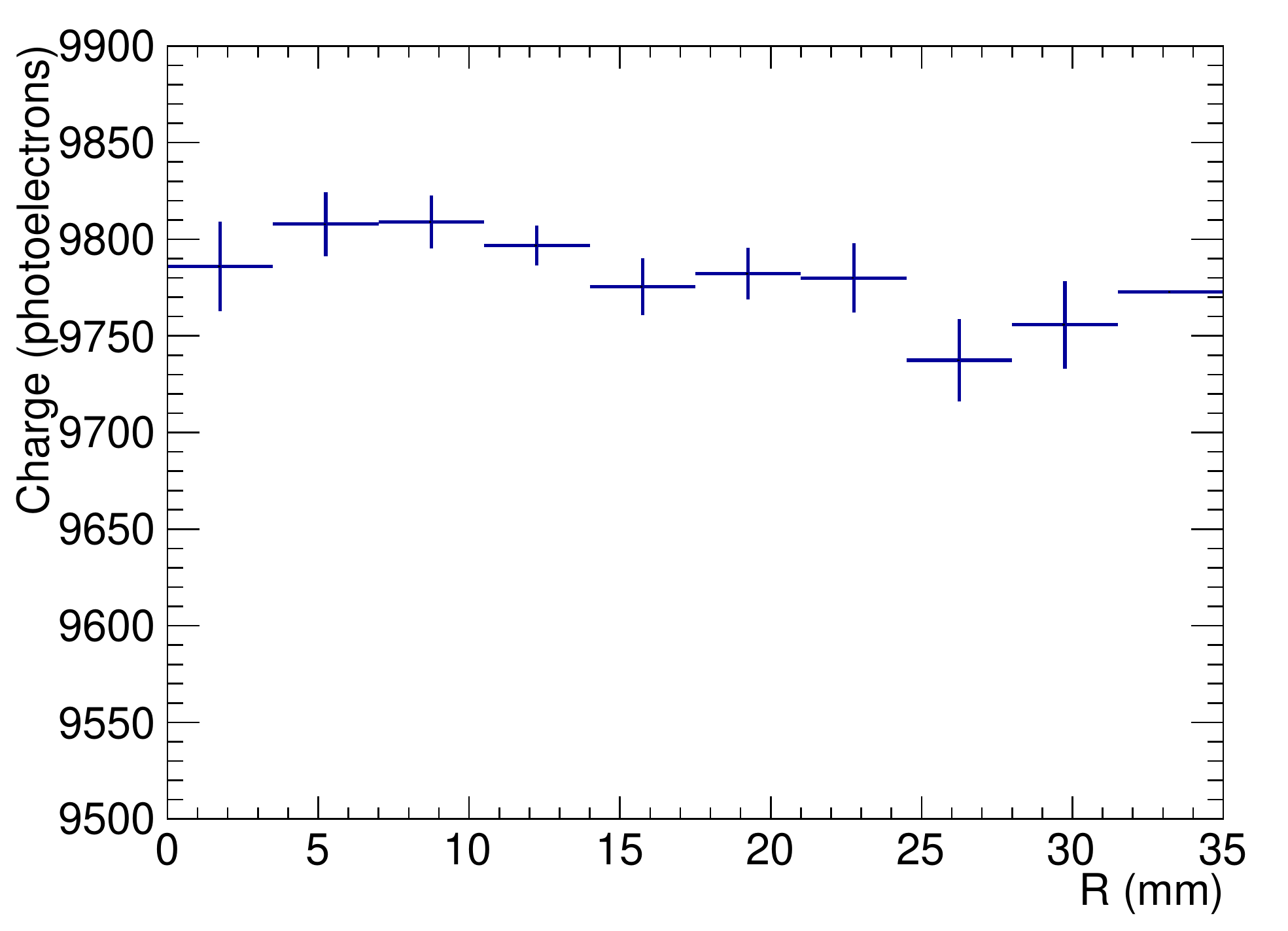}
\includegraphics[width=0.495\textwidth]{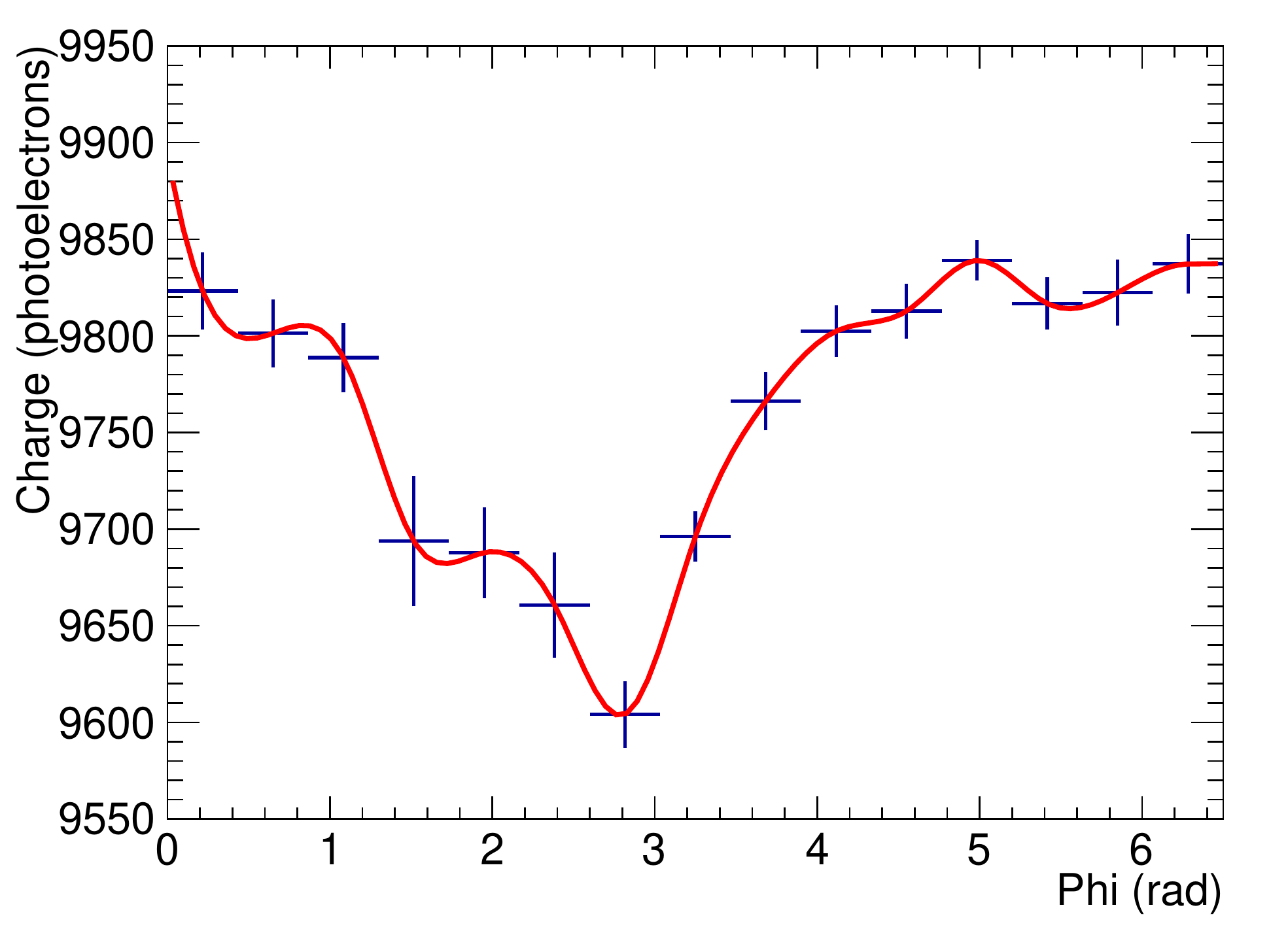}
\includegraphics[width=0.495\textwidth]{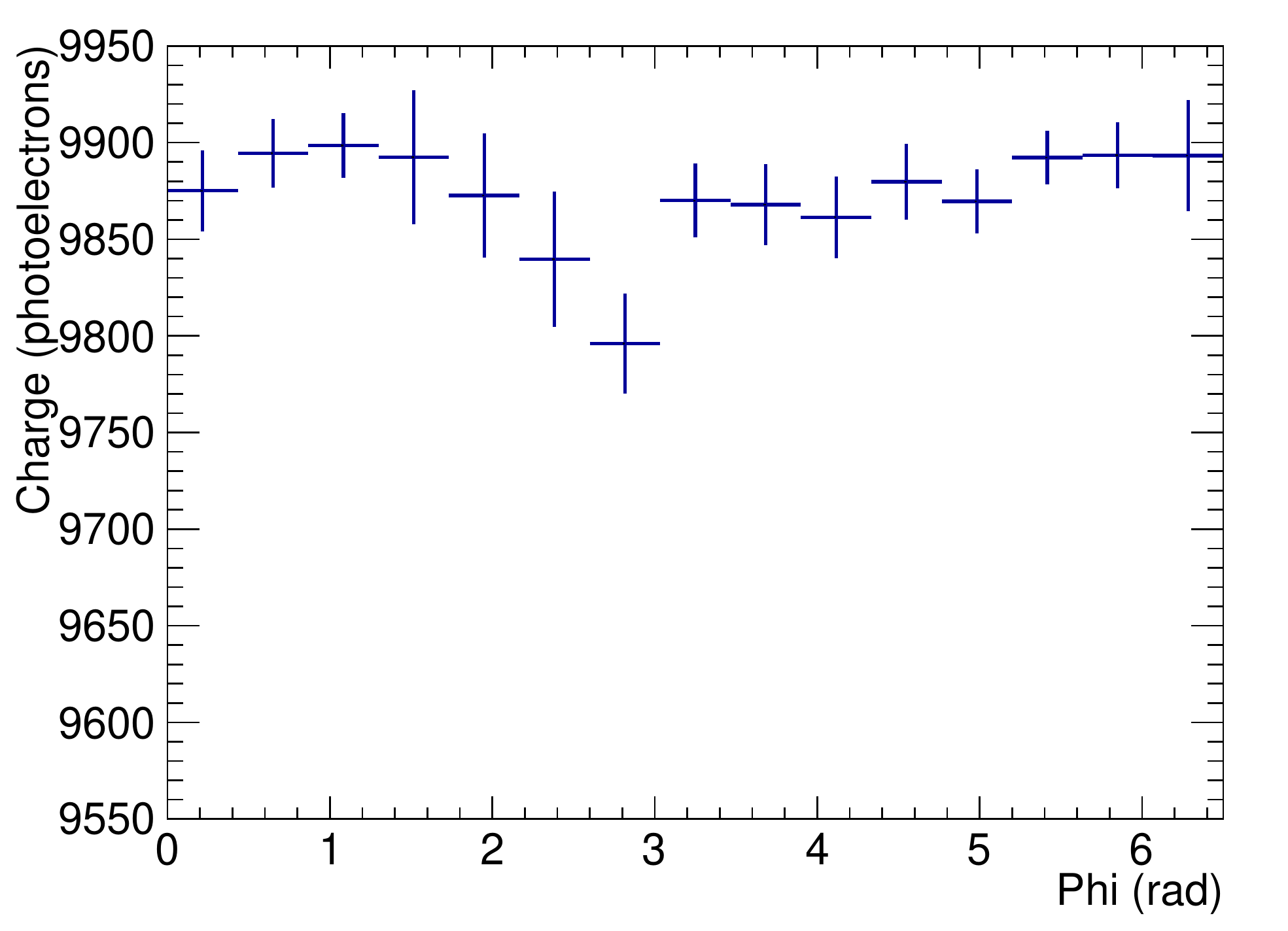}
\caption{Radial (top panels) and azimuthal (bottom panels) corrections to the signal recorded by the NEXT-DEMO energy plane under the \emph{blue} configuration. The  correction is described by a spline fitted to the data (left panels), and then applied to the data event by event (right panels).} \label{fig:RPhiCorrections}
\end{figure}
%%%%%%%%%%

The radial correction --- that is, the term $\rho(R')$ in equation~(\ref{eq:SpatialCorr}) --- was computed numerically by fitting a spline function to the profile histogram (figure~\ref{fig:RPhiCorrections}, top left panel) of the recorded energy with the reconstructed radius for photoelectric events and normalizing to the value for $R=0$. The correction was then applied event by event. The result can be seen in the top right panel of figure~\ref{fig:RPhiCorrections}. The azimuthal correction, $\Phi(\phi')$, was calculated the same way using the $\rho(R')$-corrected data --- see the bottom panels of figure~\ref{fig:RPhiCorrections}.

%%%%%%%%%%
\begin{figure}
\centering
\includegraphics[width=0.495\textwidth]{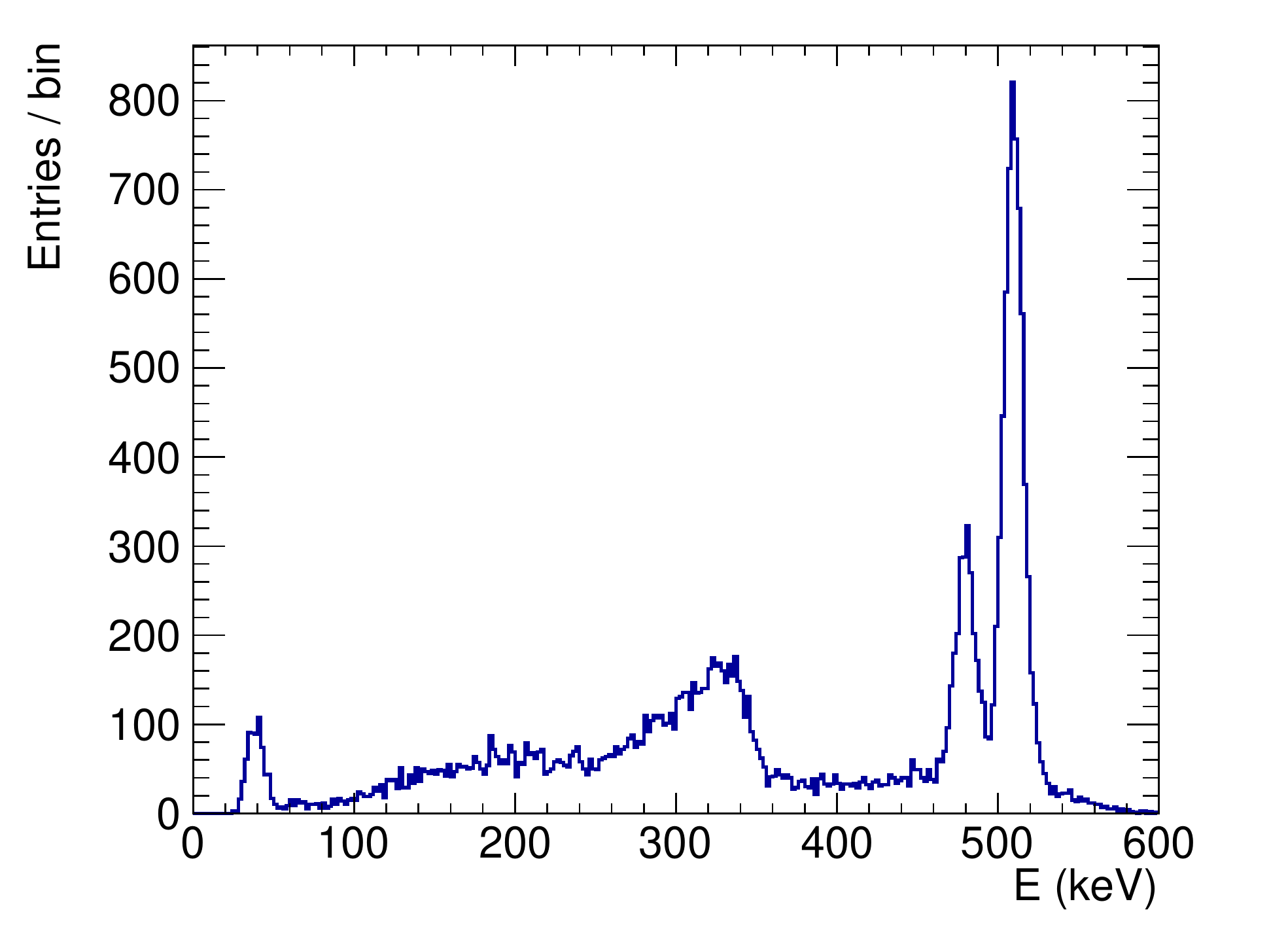}
\includegraphics[width=0.495\textwidth]{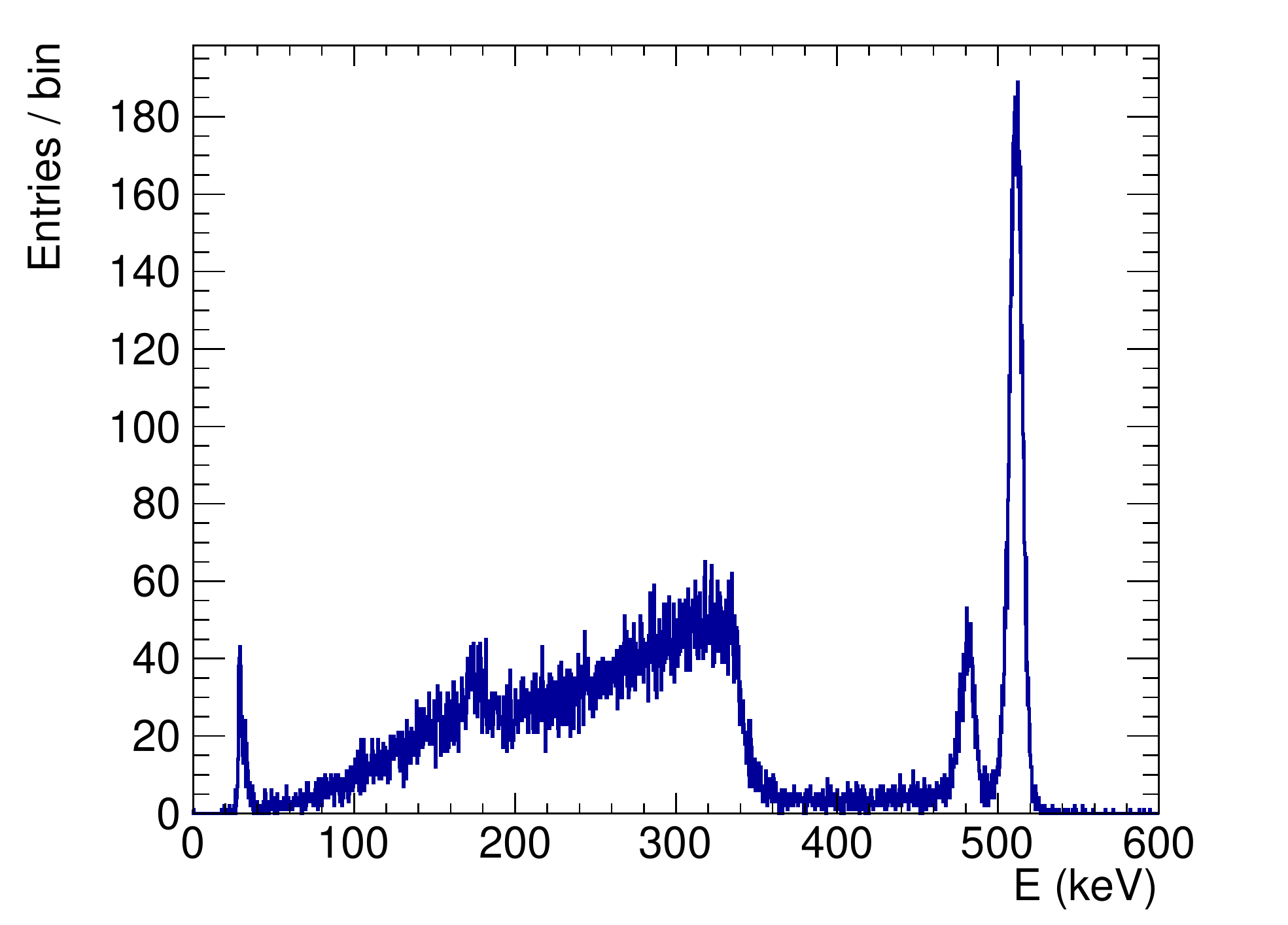}
\includegraphics[width=0.495\textwidth]{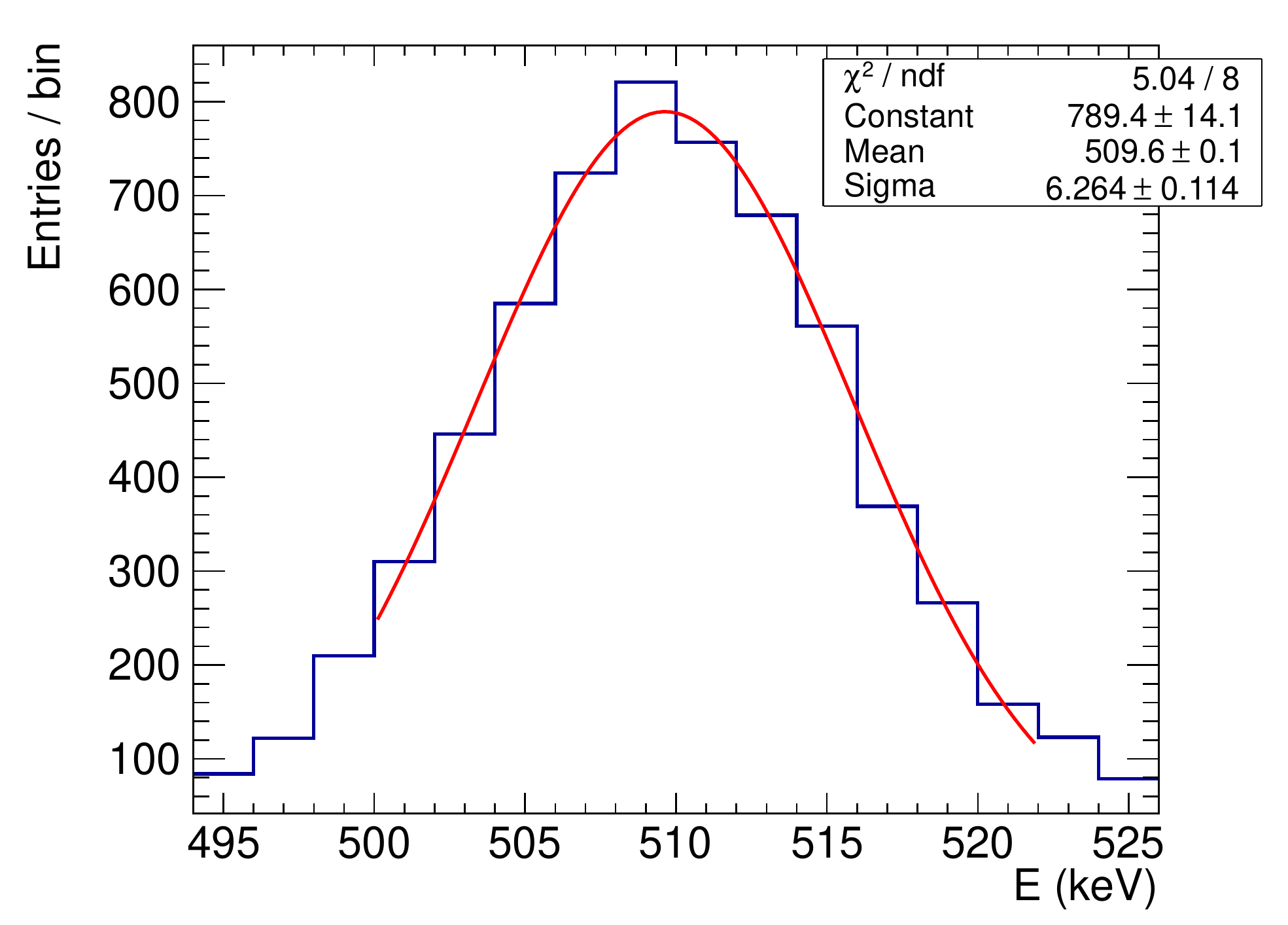}
\includegraphics[width=0.495\textwidth]{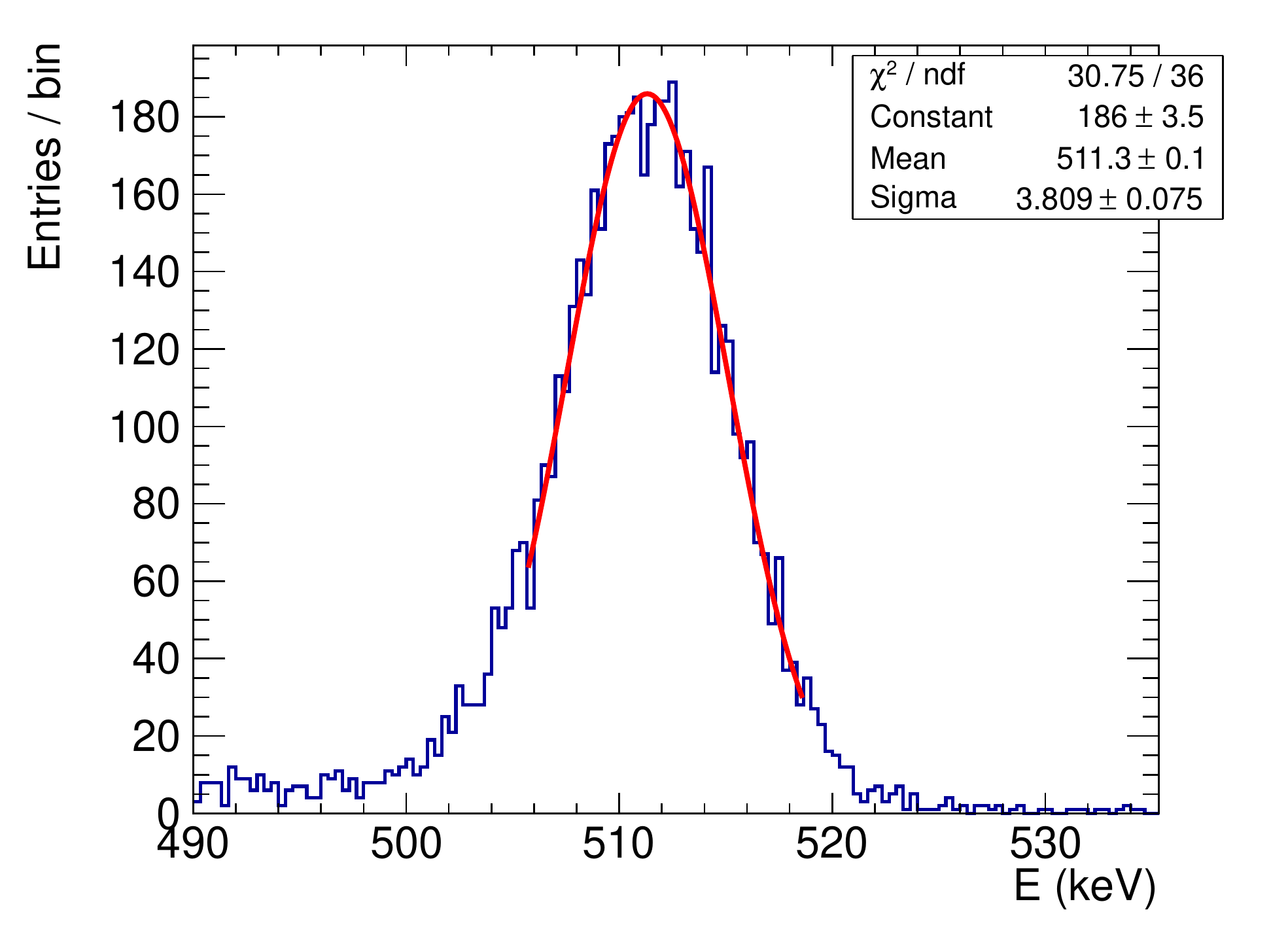}
\caption{Top: Energy spectra for $^{22}$Na gamma-ray events within the fiducial volume of NEXT-DEMO after spatial corrections. Left for the \emph{ultraviolet} configuration (UVC); right for the \emph{blue} configuration (BC). Bottom: Gaussian fits to the photoelectric peaks of the above energy spectra, indicating an energy resolution at 511 keV of 2.89\% FWHM for the UVC and 1.75\% FWHM for the BC.} \label{fig:FinalSpectrum}
\end{figure}
%%%%%%%%%%

Figure~\ref{fig:FinalSpectrum} shows the photoelectric peak region of the corrected energy spectrum of 511-keV gamma rays from \NA\ in the fiducial volume of NEXT-DEMO for the two studied configurations. A gaussian fit to the peak yields an energy resolution of 2.89\% FWHM for the UVC and 1.75\% FWHM for the BC. Improved energy resolution for BC period data is explained by the improved photoelectron statistics, which is a factor of three larger than in the UVC. Extrapolating using the BC result to the $Q$ value of \XE\ (2458~keV) assuming a $E^{-1/2}$ dependence predicts a resolution of 0.8\% FWHM. This result improves on the NEXT target resolution of 1\% FWHM at \Qbb.

%%% SECTION 6. TOPOLOGICAL SIGNATURE
%%%%%%%%%%%%%%%%%%%%%%%%%%%%%%%%%%%%%%%%%%%%%%%%%%%%%%%%%%%%
\section{The topological signature} \label{sec:Topology}
%%%
As discussed in earlier sections, NEXT will use a dedicated tracking plane to reconstruct the position and topology of any event recorded in the detector. The purpose of this analysis will be two-fold: to accurately reconstruct the position of the deposited energy so that it can be corrected for the geometrical variation in observed charge (discussed in section~\ref{sec:EnergyResolution}); and the reconstruction of the topology of an event so that background events can be efficiently rejected. The latter study will take advantage of the distinct topology left in HPXe by electrons, a short tortuous track ending in a concentrated energy deposit corresponding to the showering of the particle, hereafter referred to as a \emph{blob}. This topological signature can be used in the search for \bbonu\ to distinguish between signal (two electron tracks with a common vertex) and background (mostly single electrons originated in the interaction of high-energy gammas), as demonstrated in the Gotthard experiment~\cite{Luscher:1998sd}.

\begin{figure}
\centering
\includegraphics[height=6.1cm]{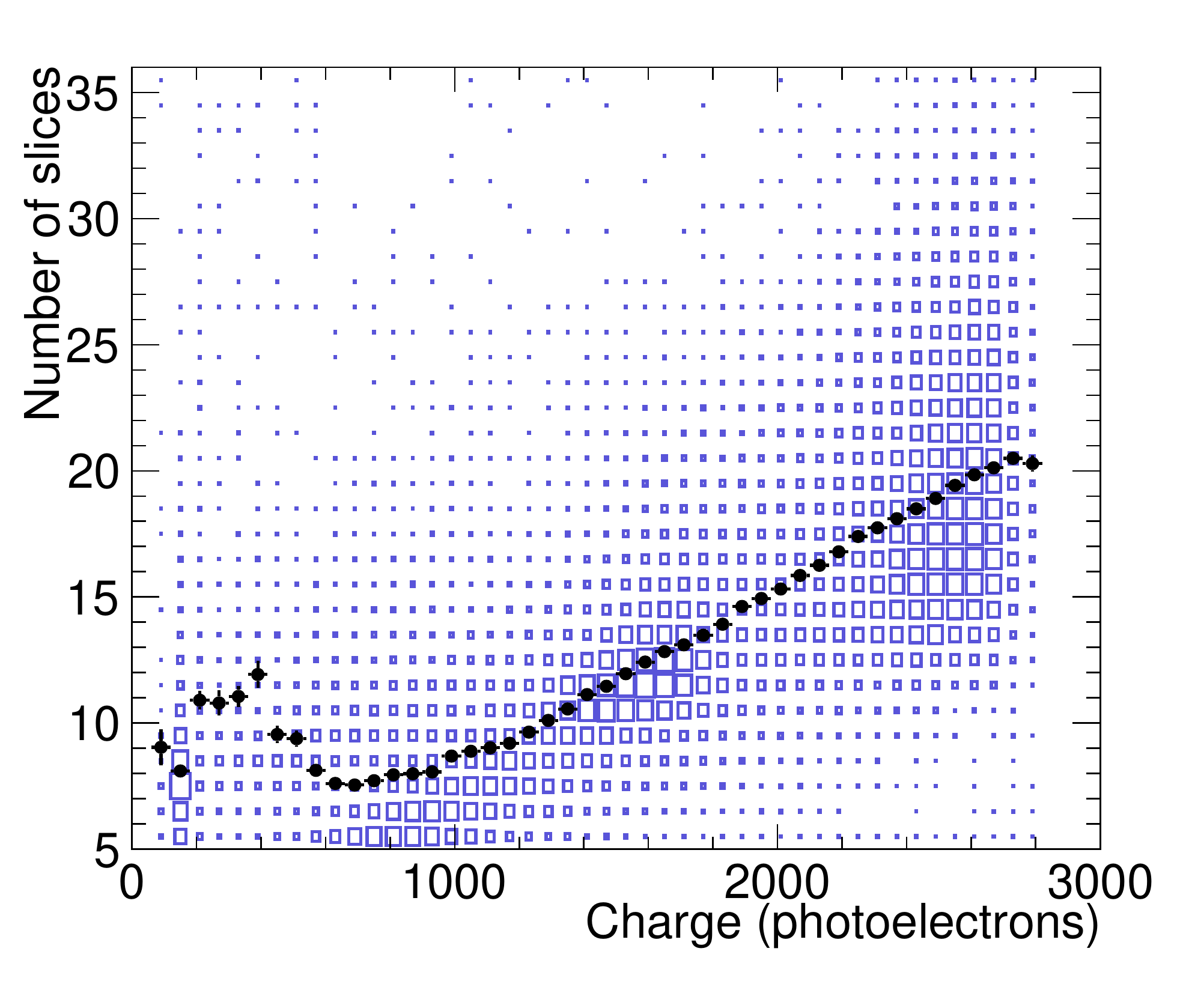}
\includegraphics[height=6.2cm]{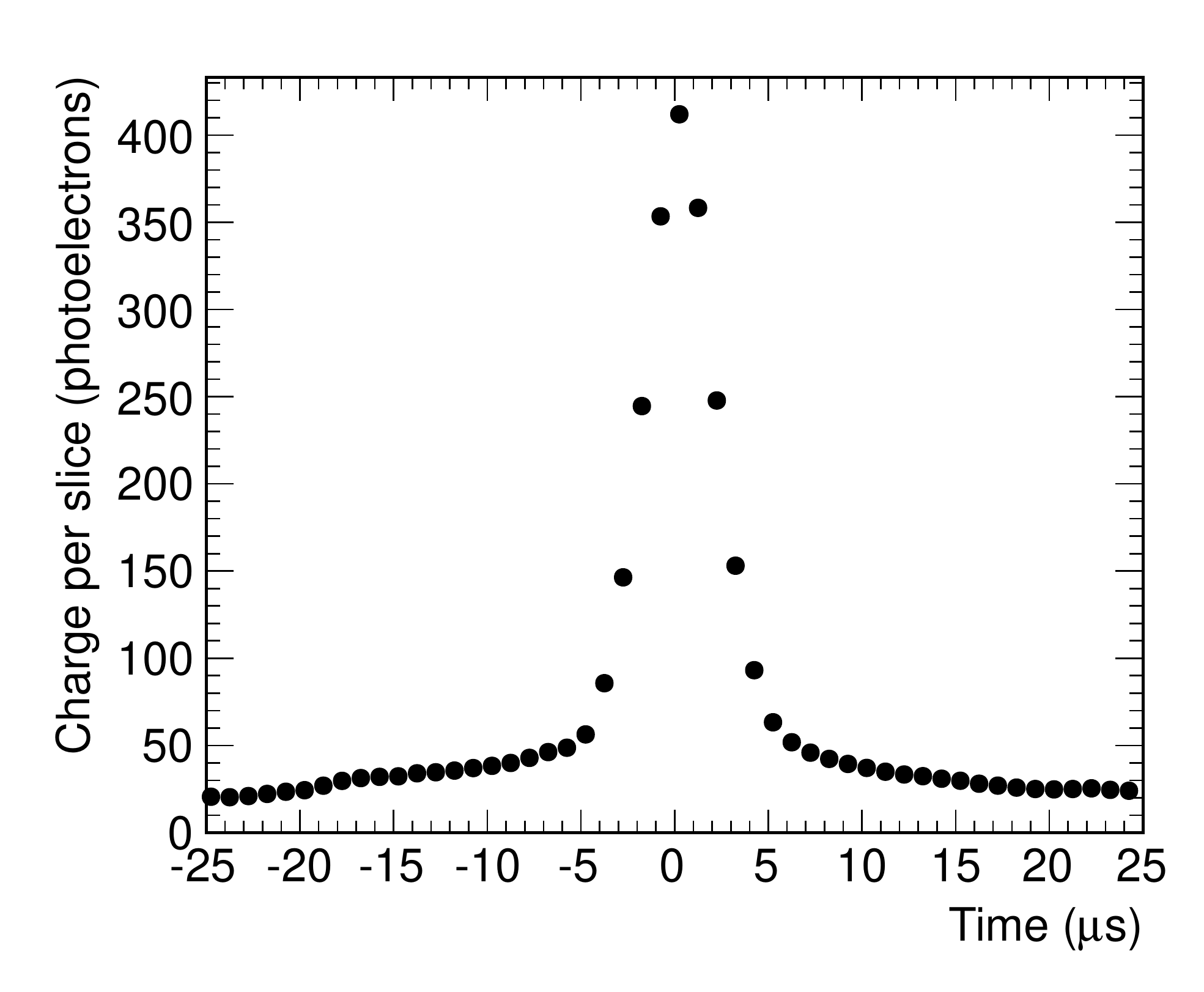} 
\caption{Left: Number of longitudinal slices in an S2 peak as a function of its integrated charge (energy). The average number (black dots) is superimposed on the 2D histogram (blue boxes). Right: Time-ordered distribution of slices with respect to the most energetic one for photoelectric events.} \label{fig:slices} 
\end{figure}

This preliminary study of tracking in NEXT-DEMO is concerned only with the $z$ coordinate since the accuracy with which this coordinate can be reconstructed is significantly higher than that of $xy$ in the current set-up. Considering the charge in individual time samples (\emph{slices}) as opposed to the full integrated charge of the selected \emph{peak} (see section~\ref{subsec:Waveforms}) makes it possible to map the $z$ evolution of an event. As can be seen in figure~\ref{fig:slices}, right panel, for the \NA\ data considered here the most energetic slice tended to be centred in the event with a sharp drop off in charge either side in $z$. This lends weight to the assumption in section~\ref{sec:EnergyResolution} that \NA\ events were essentially point-like for the purpose of energy correction. Additionally, this observation can be used to make a tentative definition of what can be classified as a \emph{blob}. That is, a slice in an event with an energy of at least 300 pes whose neighbour slices have an energy greater than 50 pes. In 98\% of the photoelectron events considered in this preliminary analysis a slice fulfilling this criterion was found, with only 0.14\% of the cases exhibiting two or more \emph{blobs}. 

The relation between the number of slices and the total charge in the peak is less simple (figure~\ref{fig:slices}, left panel) when considering only 1 dimension since an event of given energy can move in any direction within the detector. However, as is evident in the left panel of figure~\ref{fig:slices}, a peak with a greater integrated charge tends to have a greater extent in $z$. A higher energy electron has higher probability to travel a greater distance before it starts to shower since the energy loss, ${\rm d}E/{\rm d}x$, remains approximately constant in this regime and, as such, those electrons travelling parallel or at a small angle to the drift field have greater probability to be recorded over a greater number of slices.

%%%%%%%%%%
%%%%%%%%%%

%%%%%%%%%%
\begin{figure}
\centering
\includegraphics[width=0.65\textwidth]{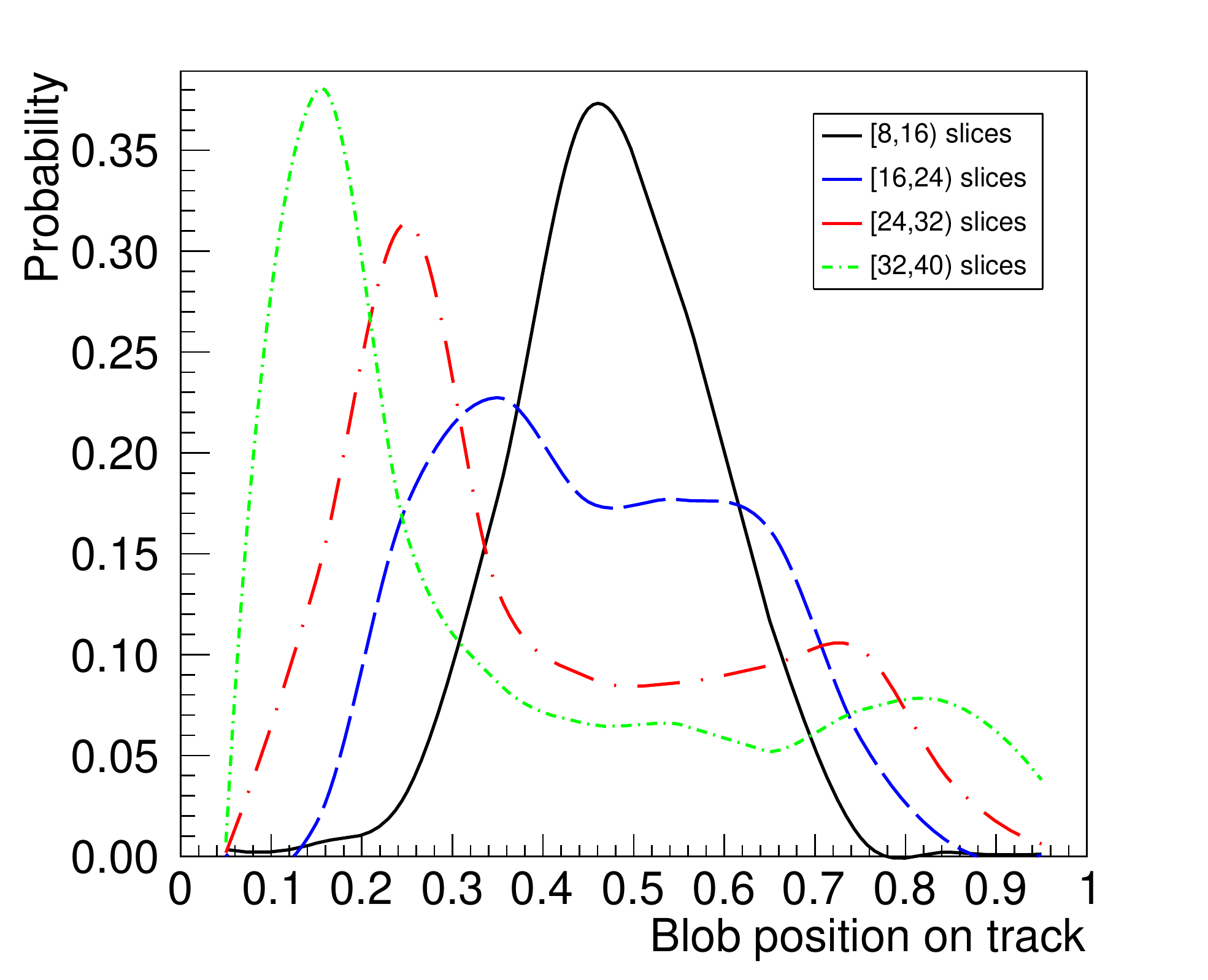} 
\caption{Relative position of the slice of maximum energy along the slices for different number of slices. Each distribution normalized to number of events considered.} \label{fig:ipos} 
\end{figure}
%%%%%%%%%%

Considering tracks with an extent greater than 7 slices in time and with total integrated charge greater than 2500 pe where a slice fulfilling the simple blob definition above is found, it can be seen in figure~\ref{fig:ipos} that this blob tends to be positioned closer to the edge of events with greater extent in $z$. This conclusion is consistent with events travelling towards the anode tending to have greater extent in $z$ and that the blob is associated with the showering of the electron at the end of its track. The bias in one direction is due to the position of the source port used next to the cathode.

%%% SECTION 7. SUMMARY AND OUTLOOK
%%%%%%%%%%%%%%%%%%%%%%%%%%%%%%%%%%%%%%%%%%%%%%%%%%%%%%%%%%%%
\section{Summary and outlook} \label{sec:Summary}
%%%
The initial results of the NEXT-DEMO detector operated with a tracking plane equipped with photomultiplier tubes and running under two configurations, with a light tube made of PTFE panels surrounding the active volume and with tetraphenyl butadiene (TPB) coated PTFE panels, have been presented. The detector response was studied with a \NA\ source. The notable increase in the recorded signals after coating ---~about a factor of 3 difference for the same energy~--- and the consequent improvement in energy resolution demonstrate the rationale behind the use of TPB. 

The energy spectrum of electrons produced by the interacting 511-keV gammas in the gas has been reconstructed in a cylindrical fiducial volume of 35 mm radius. The best resolution obtained, extrapolated to \Qbb, was 0.8\% FWHM, improving on the target defined in the experiment TDR \cite{Alvarez:2012haa} of 1\% FWHM at \Qbb\ while using only basic corrections to the directly detected signals. The measured electron lifetime was long, of the order of several milliseconds, showing that the gas recirculation through hot getters, as foreseen for NEXT-100, effectively removes the electronegative impurities from the gas. Initial studies concerning the reconstruction of the topology, in particular the identification of the electron blob (the large energy deposit associated with the end of an electron track), have also been presented. 

Future publications will employ an improved tracking plane instrumented with SiPMs \cite{Alvarez:2012za} allowing for the use of an increased fiducial region and the consideration of tracking and time dependent corrections to the energy measurement.

%%%%%%%%%%%%%%%%%%%%%%%%%%%%%%%%%%%%%%%%%%%%%%%%%%%%%%%%%%%%
\acknowledgments
This work was supported by the following agencies and institutions: the Ministerio de Econom\'ia y Competitividad of Spain under grants CONSOLIDER-Ingenio 2010 CSD2008-0037 (CUP) and FPA2009-13697-C04-04; the Director, Office of Science, Office of Basic Energy Sciences, of the US Department of Energy under contract no.\ DE-AC02-05CH11231; and the Portuguese FCT and FEDER through the program COMPETE, project PTDC/FIS/103860/2008. J.~Renner (LBNL) acknowledges the support of a US DOE NNSA Stewardship Science Graduate Fellowship under contract no.\ DE-FC52-08NA28752.

%%%%%%%%%%%%%%%%%%%%%%%%%%%%%%%%%%%%%%%%%%%%%%%%%%%%%%%%%%%%
\bibliographystyle{JHEP}
\bibliography{references}

\end{document}